\documentclass[preprint,authoryear,3p,twocolumn]{elsarticle}

\usepackage{amssymb}
\usepackage{amsmath}
\usepackage{natbib}
\usepackage{xcolor}
\usepackage{subfigure}
\usepackage{ctable}
\usepackage{makecell}
\usepackage{caption} 
\usepackage[shortlabels]{enumitem}
\usepackage{hyperref}
\usepackage{xspace}

\newcommand{\mumeaneff}{\ensuremath{\bar{\mu}_{\rm eff}}\xspace}

\usepackage{xcolor}
\usepackage{aas_macros}

\journal{Astronomy and Computing}

\begin{document}

\begin{frontmatter}



\title{DeepShadows: Separating Low Surface Brightness Galaxies from Artifacts using Deep Learning}

\author[label1,label2]{Dimitrios Tanoglidis\corref{label4}}
\cortext[label4]{Corresponding author{ead}}
\ead{dtanoglidis@uchicago.edu}
\author[label3]{Aleksandra \'Ciprijanovi\'c}
\author[label3,label2,label1]{Alex Drlica-Wagner}

\address[label1]{Department of Astronomy and Astrophysics, University of Chicago, Chicago, IL 60637, USA}
\address[label2]{Kavli Institute for Cosmological Physics, University of Chicago, Chicago, IL 60637, USA}
\address[label3]{Fermi National Accelerator Laboratory, P. O. Box 500, Batavia, IL 60510, USA}


\begin{abstract}
Searches for low-surface-brightness galaxies (LSBGs) in galaxy surveys are plagued by the presence of a large number of artifacts (e.g., objects blended in the diffuse light from stars and galaxies, Galactic cirrus, star-forming regions in the arms of spiral galaxies, etc.) that have to be rejected through time consuming visual inspection. In future surveys, which are expected to collect hundreds of petabytes of data and detect billions of objects, such an approach will not be feasible. We investigate the use of convolutional neural networks (CNNs) for the problem of separating LSBGs from artifacts in survey images. We take advantage of the fact that, for the first time, we have available a large number of labeled LSBGs and artifacts from the Dark Energy Survey, that we use to train, validate, and test a CNN model. That model, which we call \textit{DeepShadows}, achieves a test accuracy of $92.0 \%$, a significant improvement relative to feature-based machine learning models. We also study the ability to use transfer learning to adapt this model to classify objects from the deeper Hyper-Suprime-Cam survey, and we show that after the model is retrained on a very small sample from the new survey, it can reach an accuracy of $87.6\%$. These results demonstrate that CNNs offer a very promising path in the quest to study the low-surface-brightness universe.
\end{abstract}

\end{frontmatter}


\section{Introduction}
\label{sec: inroduction}

Our understanding of galaxy formation, evolution, and the relationship between galaxies and the dark matter halos that they inhabit \citep[e.g., the ``galaxy--halo connection'';][]{Wechsler_Tinker2018} is constrained by our ability to detect faint galaxies \citep[e.g.,][]{Sugata2020}.
Low-surface-brightness galaxies (LSBGs) are conventionally defined as galaxies with a central surface brightness fainter than the night sky ($\mu(g) \gtrsim 22$ mag/arcsec$^2$). 
Thus, by definition, they are very difficult to detect and characterize. 
While contributing only a small fraction to the observed luminosity of the local universe, theoretical \citep[e.g.,][]{Martin2019} and observational \citep[e.g.,][]{Dalcanton1997} arguments suggest that LSBGs account for the majority of galaxies, which thus remains relatively unexplored.

Most of the searches for LSBGs to date have targeted small regions of the sky and have revealed LSBG populations in massive galaxy clusters such as Virgo \citep[e.g.,][]{Sabatini2005,Mihos2015,Mihos2017}, Coma \citep[e.g.,][]{Adami2006,vanDokkum2015} and Fornax \citep[e.g.,][]{Hilker1999,Munoz2015,Venhola2017}, as well as faint satellites around the nearby galaxies \citep[e.g.,][]{McConnachie2012,Martin2013,Merritt2016,Danieli2017,Cohen2018}. To better understand and test galaxy formation models in the low-surface-brightness regime, it is imperative to study LSBGs over a wide sky area and across different environments (inside galaxy clusters vs.\ field). Wide-field galaxy surveys have already started to reveal a large number of LSBGs. For example, the Hyper Suprime-Cam Subaru Strategic
Program (HSC SSP)\footnote{\url{https://hsc.mtk.nao.ac.jp/ssp/}} discovered 781 radially extended (half-light radius $r_{1/2} > 2.5''$) LSBGs with $\mumeaneff(g)> 24.3$ mag/arcsec$^2$ in an analysis of the first $\sim 200$ deg$^2$ of its Wide layer \citep{Greco2018}. More recently, an analysis of the first three years of data from the Dark Energy Survey (DES)\footnote{\url{https://www.darkenergysurvey.org/}}, covering $\sim 5,000$ deg$^2$ on the southern sky, brought to light a population of $>$20,000 LSBGs with similar size and surface brightness limits \citep{Tanoglidis2020}. 

Searches for LSBGs in survey data are plagued by the presence of a large number of low-surface-brightness artifacts in astronomical images. Examples of these artifacts include:
\begin{itemize}
    \item Faint, compact objects blended in the diffuse light from nearby bright stars or giant elliptical galaxies;
    \item Bright regions of Galactic cirrus;
    \item Knots and star-forming regions in the arms of large
spiral galaxies;
    \item Tidal ejecta connected to high-surface-brightness
host galaxies.
\end{itemize}
Such objects often pass the selection criteria outlined above and dominate the sample of candidate LSBGs.
For example, in DES there were 413,000 LSBG candidates, with only ${\sim}5\%$ of them being genuine LSBGs. Even after a feature-based machine learning (ML) classification step that reduced the sample by approximately an order of magnitude, a large number of false-positives remained (${\sim}50\%$ of objects classified as LSBGs). These false-positives had to be manually rejected through visual inspection. Similarly, the authors of the HSC SSP study had to go through a visual inspection step, since their pipeline produced a sample that also had a ${\sim}50\%$ contamination rate from artifacts \citep{Greco2018}. 

Visual inspection is time consuming and difficult to perform systematically. Upcoming galaxy surveys, such as the Legacy Survey of Space and Time (LSST)\footnote{\url{https://www.lsst.org/}} on the Vera C.\ Rubin Observatory\footnote{\url{https://www.vro.org/}} and Euclid\footnote{\url{https://www.euclid-ec.org/}} are expected to produce massive volumes of data. 
LSST will observe $\sim 20{,}000$ deg$^2$ of sky, produce 20TB of data per night, and observe ${\sim} 10$ billion galaxies over its 10 years of observations.\footnote{\url{https://www.lsst.org/scientists/keynumbers}} With such volumes of data, rejecting artifacts via visual inspection will be impossible. Clearly, the process has to be automated.

Machine learning, and in more recent years deep learning, have started to revolutionize astronomy as the sizes of astronomical datasets grow \citep[for reviews see e.g.,][]{Ball2010,Baron2019}. Classification tasks are one of the classical examples where machine learning techniques can be applied. In cases dealing with high-dimensional feature spaces where large training sets are available, deep learning usually outperforms other machine learning algorithms and reaches human-level performance \citep{lecun2015deeplearning}.

Convolutional neural networks \citep[CNNs;][]{LeCun1998} constitute a specific class of deep learning algorithms, inspired by the visual cortex and optimized for computer vision tasks. For that reason they are a promising tool for analyzing astronomical images. Furthermore, working directly at the image level (with the pixels as inputs) eliminates the need for deriving and selecting parameters (sizes, magnitudes, colors etc.) as features to be used for the classification task, which can be subjective and non-optimal.

CNNs were first introduced in astronomy by \citet{Dieleman2015} to perform automatic morphological classification of galaxies and have since found a number of applications. For example, other authors further explored their use in classifying galaxy morphologies \citep[e.g.,][]{Dai2018,Sanchez2018,Cheng2019}, separating stars from galaxies \citep[e.g.,][]{Kim2016}, identifying strong lenses \citep[e.g.,][]{Lanusse2018,Jacobs2019,Davies2019,Bom2019}, eliminating polarimetric artifacts \citep{Paranjpye2020}, evaluating flare statistics in young stars \citep{Feinstein2020}, classifying galaxy mergers \citep{Ciprijanovic2020}, reconstructing lensing of the Cosmic Microwave Background \citep{Caldeira2018}, setting constraints on the cosmological parameters from weak lensing \citep[e.g.,][]{Ribli2019}, and many other applications.

In this paper we present the first application of CNNs in classifying LSBGs in astronomical images and we demonstrate that they can significantly help in automating this process. We  take advantage of the fact that for the first time we have available a large sample of LSBGs and an equally large number of labeled artifacts from visual inspection \citep{Tanoglidis2020}. These large labeled training sets are necessary to successfully train CNN models. We compare the performance of our CNN architecture to conventional machine learning models (support vector machines and random forests) trained on features extracted from the same objects. We also study how well the model trained on the DES images can classify images from the HSC SSP, thus demonstrating promise for using a similar technique in upcoming surveys, such as LSST.

This paper is organized as follows: In \S \ref{sec: Data} we describe the datasets we use for the classification problem. In \S \ref{sec: Methods} we briefly summarize the theory and formalism of neural networks and present the specific architecture we use to tackle the problem at hand, which we call \textit{DeepShadows}. In \S \ref{sec: Results} we present the classification results. In \S \ref{sec: Transfer_learning} we use transfer learning to classify objects from the HSC SSP for which we have labels. In \S \ref{sec: Errors} we study the uncertainties present in our study and their impact on the metrics used to assess the classification performance of the model. We discuss our results, propose paths for future investigation, and conclude in \S \ref{sec: Discussion_Conclusions}.

The code and data related to this work are publicly available at the GitHub page of this project: {\url{https://github.com/dtanoglidis/DeepShadows}}.
 
\section{Data}
\label{sec: Data}

\begin{figure*}[!ht]
\centering
\subfigure[]{\includegraphics[width=0.48\textwidth]{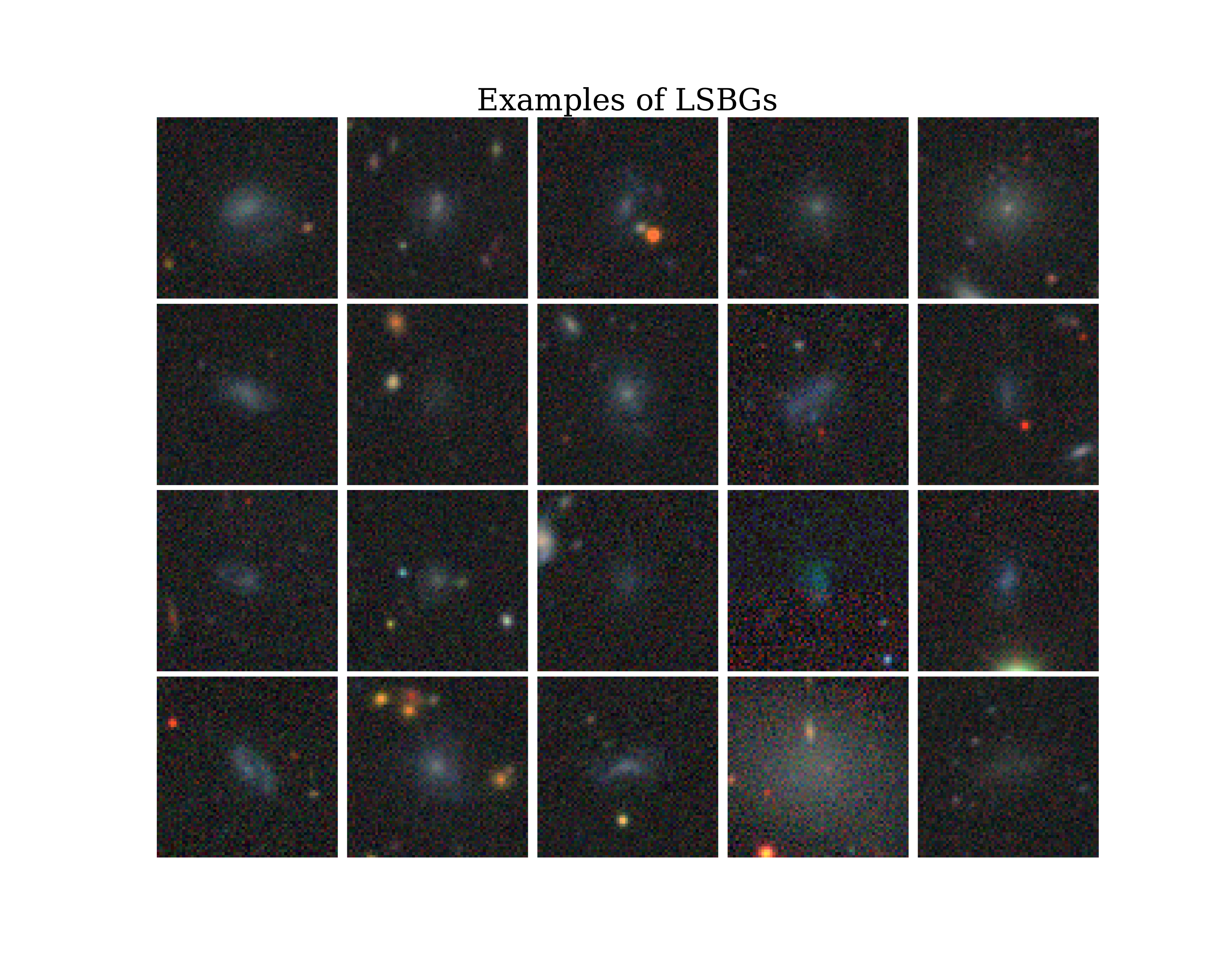}}
\hspace*{\fill}
\subfigure[]{\includegraphics[width=0.48\textwidth]{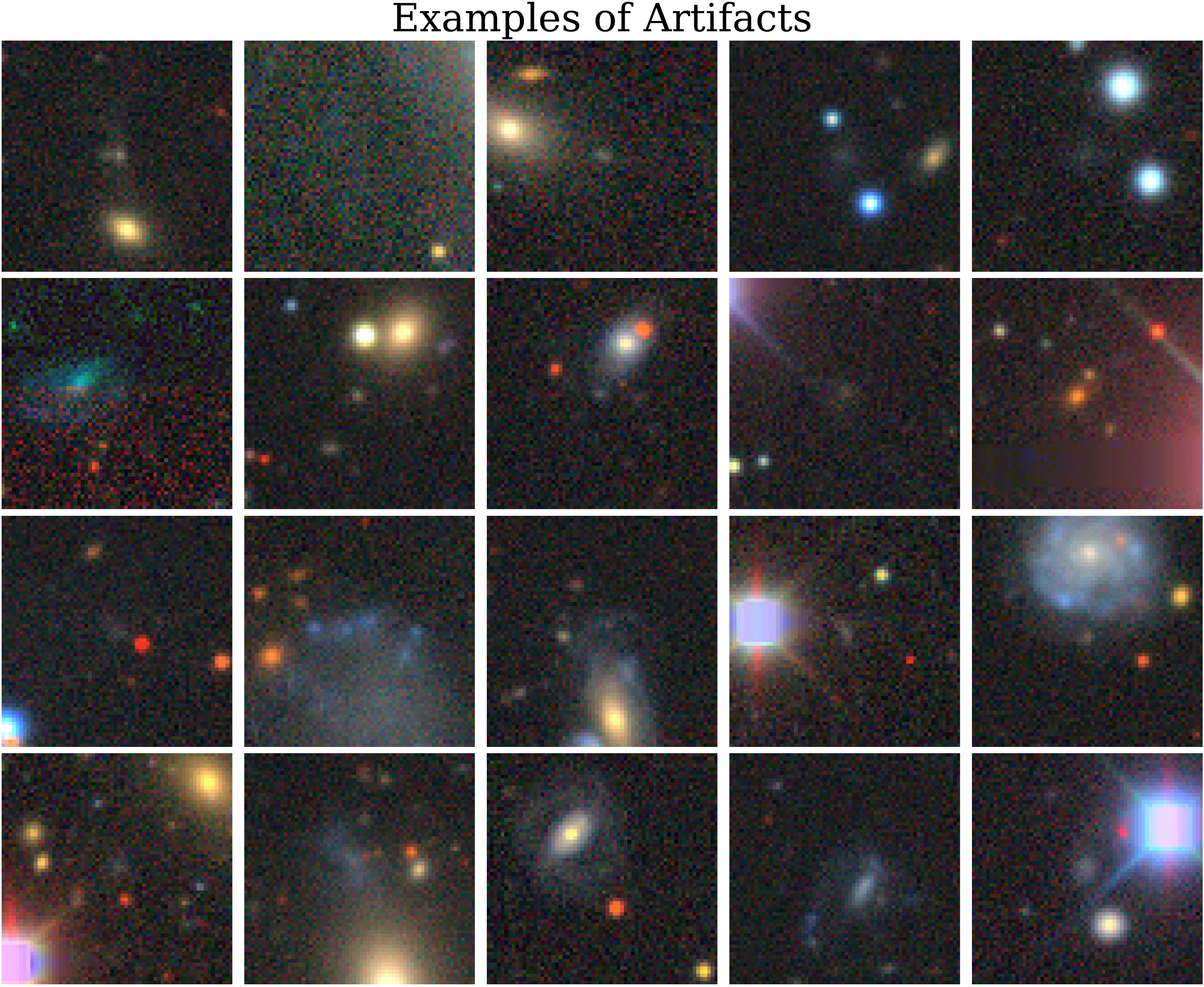}}
\caption{Example images of (a) LSBGs  and (b) artifacts in our dataset. Each cutout corresponds to a $30'' \times 30''$ angular region on the sky. }
\label{fig:Cutout_Examples}
\end{figure*}

In this section we describe the datasets used for training and evaluating the performance of the CNN and other machine learning models. We briefly describe the astronomical surveys and selection procedures used to obtain these data.

\subsection{The Dark Energy Survey}

Our primary dataset comes from the first three years (Y3) of observations by DES. DES is an optical/near infrared imaging survey that covers ${\sim} 5,000$ deg$^2$ of the southern Galactic cap in five photometric filters, $grizY$, to a depth of $i \sim 24$ over the course of a six-year observational program with the 570-megapixel 
Dark Energy Camera (DECam) on the 4m Blanco Telescope at the
Cerro Tololo Inter-American Observatory (CTIO) in Chile. The DECam field-of-view covers 3 deg$^2$ with a central
pixel scale of $0.263''$ \citep{Flaugher2015}. 

Objects were detected in astronomical images using \texttt{SourceExtractor} \citep{Bertin_Arnouts1996}, which provides a catalog of photometric parameters for each object, such as magnitudes, flux radii, mean and central surface brightnesses (adaptive aperture measurements) etc. For a more detailed description of the DES data, see \citet{Abbott2018}.

\subsection{LSBGs and artifacts in DES}
\label{sec: LSB_and_Art}

We train, validate and test the performance of our models on LSBGs and artifacts detected by DES, as described in \citet{Tanoglidis2020}. Here, we briefly outline the main steps followed in that paper for the LSBG catalog construction: 
\begin{enumerate}
\item Selection cuts were performed using the \texttt{SourceExtractor} parameters from the full DES catalog. The most important cuts are on the angular size (half-light radius in the $g$-band, $r_{1/2}>2.5''$) and on the mean surface brightness within the effective radius
($\mumeaneff(g)>24.3$ mag/arcsec$^2$).\footnote{An updated version of \citet{Tanoglidis2020} uses a brighter selection, $\mumeaneff(g)>24.2$ mag/arcsec$^2$. However, we keep the older definition, since the LSBG/artifact separation is more challenging in the fainter regime.}
\item Classification was performed using Support Vector Machines (SVMs) trained on \texttt{SourceExtractor} output parameters (features) and a manually annotated set of $\sim 8{,}000$ objects, out of which 640 were LSBGs.
\item Visual inspection was used to reject false positives from more than 40,000 objects positively classified in the previous step.
\item Sérsic model fitting and interstellar extinction correction was applied to objects passing the visual inspection, and new selection cuts were performed on the updated parameters. 
\end{enumerate}
For the current classification study, we randomly select 20,000 LSBGs from those visually verified in Step 3 to be used as the positive class. In the main body of our paper we use as the negative sample $20{,}000$ objects from those visually rejected in Step 3. These are the most challenging artifacts to be separated from LSBGs, since they passed the feature-based classification step in Step 2. We consider a three-class classification problem in \ref{sec: three_class}, where we add another class of artifacts, $20{,}000$ randomly selected objects from those rejected in Step 2. 

\subsection{Generation of datasets}
\label{sec: datasets}

For the LSBGs and artifacts we consider two datasets: parameters from \texttt{SourceExtractor} to be used as features for the classical machine learning models (SVMs and random forests) and images to be used in our \textit{DeepShadows} CNN model.
The features selected are:
\begin{itemize}
 \item Ellipticity of the detected objects.
\item {\texttt{MAG\_AUTO}} magnitudes in the three bands, $g, r,i$.
\item Colors $g-i$, $g-r$, $r-i$.
\item Mean, central and effective surface brightnesses in the three bands, $g, r$ and $i$.
\end{itemize}
We generate the image cutouts using the DESI Legacy Imaging Surveys Sky Viewer \citep{Dey:2019}\footnote{\url{http://legacysurvey.org/}} to access the DES DR1 (Data Release 1) images. Each image corresponds to a $30'' \times 30''$ region on the sky and is centered at the coordinates of the candidate object (LSBG or artifact). The initial size of each image is $256 \times 256$ pixels that we resize to $64 \times 64$ pixels to reduce the dataset size and the memory needs for its processing. The images also have inputs in the three RGB channels (which correspond to $g,r,z$ astronomical bands), so their size is finally is $64\times 64\times3$ (we follow a ``channels last'' format). The code we used to generate the cutouts can be found in the GitHub page of the project. In Fig.~\ref{fig:Cutout_Examples} we show examples of cutouts of LSBGs and artifacts.

Before training, we split our full sample of 40,000 objects into a training set of 30,000 examples, a validation set of 5,000 examples and a test set of 5,000 examples. We split the datasets of \texttt{SourceExtractor} parameters and images in the same way (same objects in each set).

\subsection{HSC SSP dataset}
\label{sec: HSC}

We also consider a dataset of 640 LSBGs and 640 artifacts discovered in the HSC SSP as described in \citet{Greco2018}. This is an independent set from a survey with different specifications and different human biases (in the labeling of LSBGs/artifacts) that can be used to test the ability of our model trained on the DES images to classify LSBG candidates in other surveys. We generate image cutouts for HSC SSP, using the Sky Viewer used for the DES images.
We split the full dataset of 1,280 objects into a small set of 320 objects to be used for re-training of the classifier (transfer learning) and one of 960 objects for testing the performance.

\section{Methods}
\label{sec: Methods}

In this section we introduce the notation and briefly describe the machine learning models, the neural networks, and the specific CNN architecture that we use. Our discussion is parsimonious. For a more detailed discussion we suggest the classic book by \citet{Hastie} as well as that by  \citet{Ivezic} that discusses machine learning with a focus on astrophysical applications. The book by \citet{Goodfellow} has become a standard reference for deep learning.

\subsection{Machine Learning}
\label{sec: Machine_Learning}

We consider two machine learning classification algorithms that have been proven to be powerful in a number of astrophysical problems (see the review papers mentioned in the introduction and references therein), namely SVMs and random forests. These algorithms perform best with \textit{structured} data (i.e., features/properties of the objects under study) and we apply them to the \texttt{SourceExtractor} output properties. 

Consider a number of examples, $N$, each described by a feature vector $\mathbf{x}_i$, $i = 1,\dots, N$ and with labels $y^i$ (they usually are denoted as $\{1,-1\}$ or $\{1,0\}$ in two-class problems).

The SVM classifier \citep{cortes1995support} seeks to find a separating hyperplane of the form $\mathbf{w}\cdot \mathbf{x}-b=0$, with the weights $\mathbf{w}$ and the bias $b$ selected to maximize the margin (distance) between this hyperplane and the training samples that are closest to it (support vectors). In other words, the problem can be characterized as trying to minimize the function $\frac{1}{2}||\mathbf{w}||^2$ such as $y^i(\mathbf{w} \cdot \mathbf{x}_i +b) \geq 1$ for every $i$. This formulation does not allow for misclassification (hard-margin) though, and thus is prone to overfitting. In practice, we leave some room for misclassification by using a soft-margin SVM, by trying to minimize the following function:
\begin{equation}
    \frac{1}{2}||\mathbf{x}||^2 + C \left(\sum_i^N \xi^i\right),
\end{equation}
subject to $y^i(\mathbf{w} \cdot \mathbf{x}_i +b) \geq 1 - \xi^i\,\, \forall i$. The $\xi^i$ are called slack variables and allow for misclassifications. The variable $C$ controls the ``softness" (how much error is allowed) of the classfier and is one of the tunable parameters (\textit{hyperparameters}) of the model. 

The above description assumes that the different classes are linearly separable by a hyperplane, though this is not always true. In such cases a kernel function is introduced to perform a non-linear transformation that maps the data to a space where they are linearly separable: $(\mathbf{x},\mathbf{x}') \to K(\mathbf{x},\mathbf{x}')$. A popular kernel, and the one used in this work, is the Gaussian or Radial Basis Function (RBF) kernel \citep{Orr1995}: $K(\mathbf{x},\mathbf{x}') = \exp \left(-\gamma ||\mathbf{x} - \mathbf{x}'||^2 \right)$, where $\gamma=1/2\sigma^2$ is another tunable hyperparameter.

Random forests is another powerful classification method. Random Forests \citep{Ho1995,breiman2001random} are an ensemble classifier, in the sense that use a collection of simple classifiers, known as decision trees (DTs), and the output is the majority class derived from them. 

A DT divides the dataset into subsets by setting splitting criteria on the features, trying to make these subsets as homogeneous (with respect to the labels of the examples in them) as possible. A DT starts by splitting the data on the feature that results in maximum information gain (imagine asking the most informative question when trying to make a decision) and continues until pure final subsets (nodes) are produced.

The random forests classifier considers a set of $n$ (hyperparameter) DTs and for each one uses a random selection of $\sqrt{m}$ features ($m$ is the total number of features) to construct a DT and train it on a randomly selected subset of the training examples. As mentioned above, the final result is the majority vote (class output) from these $n$ trees. 

The values of the hyperparameters are selected (tuned) by exploring a grid of possible values, and obtaining those that give the best results either by evaluating on the validation set, or using $k-$fold cross validation, where the training set is split in $k$ parts, with $k-1$ used for training and the other for validation, and then repeating this process $k$ times.

\subsection{Deep Learning}
\label{sec: Deep_Learning}

\begin{figure*}[!ht]
\centering\includegraphics[width=1.0\textwidth]{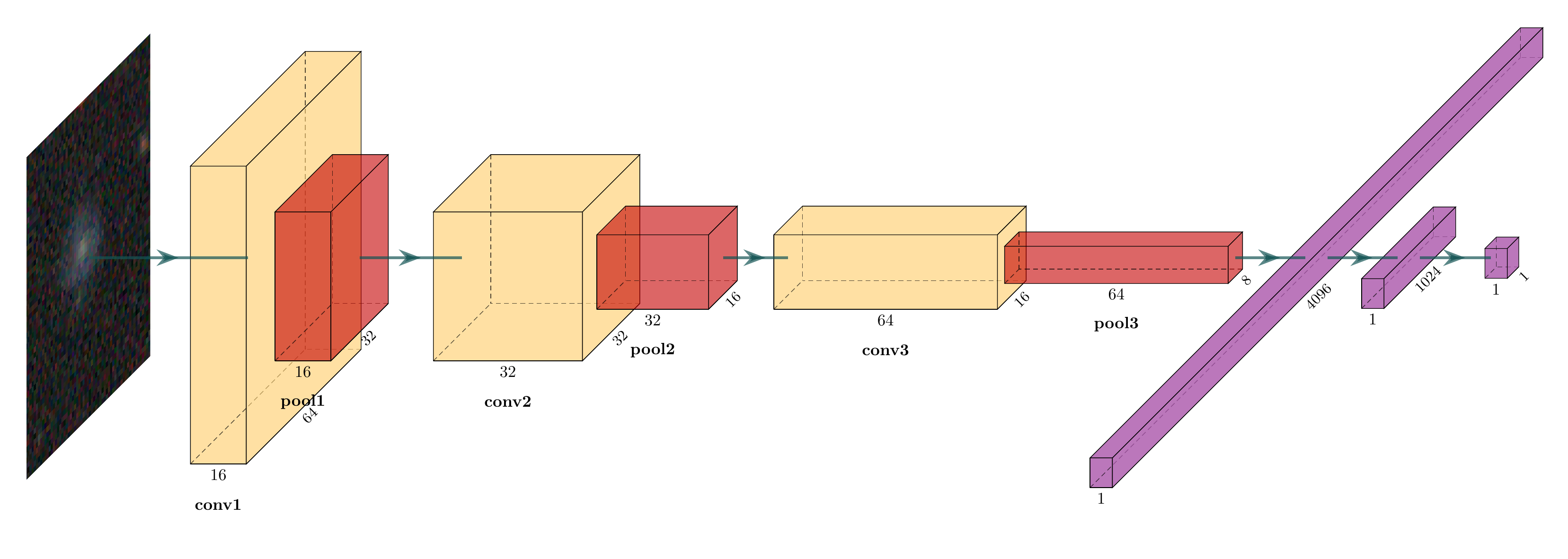}
\caption{Schematic overview of the \textit{DeepShadows} CNN architecture. There are three convolutional layers (yellow), each followed by a max pooling layer (red). The number of filters are 16, 32 and 64 for each layer, respectively. Two dense layers (purple) follow the last pooling layer after flattening. The output is a probability that the image contains an LSBG. Figure was created using the PlotNeuralNet code \citep{Iqbal2018}.}
\label{Fig: Architecture}
\end{figure*}

\begin{table*}[!ht]
  \centering
  \noindent\begin{minipage}[b]{0.99\textwidth}
   \centering
    \caption{Architecture of the {\it DeepShadows} CNN.}
  \label{table:arch}
  \centering
  \begin{tabular}{|l | l l l  l   l |}
\hline
\bf{Layers}         & \bf{Properties}           & \bf{Stride}       & \bf{Padding}  & \bf{Output Shape} & \bf{Parameters}   \\ \hline\hline
Input               & $64\times64\times3$\footnote{We use ``channel last" image data format.}       & -                 & -             & (64, 64, 3)       & 0                 \\ \specialrule{.2em}{.1em}{.1em}
Convolution (2D)    & Filters: 16                & $1\times1$        & Same          & (64, 64, 16)       & 448               \\
                    & Kernel: $ 3\times3$       & -                 & -             & -                 & -                 \\
                    & Activation: ReLU          & -                 & -             & -                 & -                 \\
                    & Reg: L2 (0.13)          & -                 & -             & -                 & -                 \\                                    \hline
Batch Normalization & -                         & -                 & -             & (64, 64, 16)       & 64               \\ \hline
MaxPooling          & Kernel: $2\times2$        & $2\times2$        & Valid         & (32, 32, 16)       & 0                 \\ \hline
Dropout             & Rate: $0.4 $              & -                 & -             & (32, 32, 16)       & 0                 \\ \specialrule{.2em}{.1em}{.1em}
Convolution (2D)    & Filters: 32               & $1\times1$        & Same          & (32, 32, 32)      & 4640              \\
                    & Kernel: $ 3\times3$       & -                 & -             & -                 & -                 \\
                    & Activation: ReLU          & -                 & -             & -                 & -                 \\ 
                    & Reg: L2 (0.13)          & -                 & -             & -                 & -                 \\                       
                    \hline
Batch Normalization & -                         & -                 & -             & (32, 32, 32)      & 128               \\ \hline
MaxPooling          & Kernel: $2\times2$        & $2\times2$        & Valid         & (16, 16, 32)      & 0                 \\ \hline
Dropout             & Rate: $0.4 $              & -                 & -             & (16, 16, 32)      & 0                 \\ \specialrule{.2em}{.1em}{.1em}
Convolution (2D)    & Filters: 64               & $1\times1$        & Same          & (16, 16, 64)      & 18496              \\
                    & Kernel: $ 3\times3$       & -                 & -             & -                 & -                 \\
                    & Activation: ReLU          & -                 & -             & -                 & -                 \\ 
                    & Reg: L2 (0.13)          & -                 & -             & -                 & -                 \\   
                                          
                    \hline
Batch Normalization & -                         & -                 & -             & (16, 16, 64)      & 256                \\ \hline
MaxPooling          & Kernel: $2\times2$        & $2\times2$        & Valid         & (8, 8, 64)        & 0                 \\ \hline
Dropout             & Rate: $0.4 $              & -                 & -             & (8, 8, 64)        & 0                 \\ \specialrule{.2em}{.1em}{.1em}
Flatten             & -                         & -                 & -             & (4096)            & -                 \\ \hline
Fully connected     & Activation: ReLU          & -                 & -             & (1024)              & 4195328            \\
                    & Reg: L2 (0.12)       & -                 & -             & -                 & -                 \\ \hline
Fully connected     & Activation: Sigmoid       & -                 & -             & (1)               & 1025                \\ \hline
\end{tabular}
\end{minipage}
\end{table*}

The standard deep learning architecture is that of a multi-layer neural network, with the outputs of the previous layer being the inputs to the following one. For example, for the $n$-th layer:
\begin{equation}
\label{eq: fully_connected}
\mathbf{x}_n = g(\mathbf{W}_n \mathbf{x}_{n-1} + \mathbf{b}_n),
\end{equation}
where $\mathbf{W}_n$ a matrix containing the weights of all the neurons of that layer and $\mathbf{b}_n$ a vector of biases. The function $g$ is called \textit{activation function} and its purpose is to introduce non-linearities in the network. 
We use the rectified linear unit (ReLU), $g(x)=\mbox{max}(0,x)$, activation function except in the final layer, where the activation function takes a sigmoid form (for a binary problem) to allow the output to be interpreted as probability.

When training a deep learning model, we seek to find a set of weights and biases that minimize a loss function between the predicted and true labels. For binary problems this is usually the cross-entropy: 
\begin{equation}
{\cal{L}}(\mathbf{y},\mathbf{\hat{y}}) = -\frac{1}{N}\sum_{i=1}^N \hat{y}^i \log_2 y^i+(1-\hat{y}^i)\log_2(1-y^i).
\end{equation}
The parameters are updated attractively via gradient descent:
\begin{gather}
\mathbf{W}_n \leftarrow \mathbf{W}_n - \eta \frac{\partial {\cal{L}}}{\partial \mathbf{W}}_n\\
\mathbf{b}_n \leftarrow \mathbf{b}_n - \eta \frac{\partial {\cal{L}}}{\partial \mathbf{b}}_n,
\end{gather}
where $\eta$ $(>0)$ is known as the \textit{learning rate} that controls the magnitude of the update at each step. In practice, only a small sample (mini-batch) is used to update the weights at each step as a way to speed up computation.

CNNs are modified versions of the network described in Eq. \ref{eq: fully_connected} inspired by the visual cortex. The main difference is that each neuron in a convolutional layer is connected only to neurons within a small rectangle in the previous layer, usually $3\times3$ to $5\times5$ pixels in size.  The output of such a layer is a predefined number of \textit{feature maps}, generated by convolving the feature maps of each previous layer with different \textit{filters} (or \textit{kernels}), whose trainable weights can capture abstract visual features.

If we have $k=1,\dots,K$ input feature maps and $\ell = 1,\dots, L$ output feature maps, in analogy to Eq. \eqref{eq: fully_connected} we write the $\ell$-th output map of the $n$-th layer as:
\begin{equation}
\label{eq: convolutional}
\mathbf{x}_n^{\ell} = g\left(\sum_{k=1} \textbf{W}_n^{k,\ell} * \mathbf{x}_{n-1}^k + b_n^\ell \right),
\end{equation}
where $*$ represents the convolution operation.

Convolutional layers are almost always followed by \textit{pooling} layers whose purpose is to subsample the output of the convolutional layer, reducing the number of trainable parameters. They have no weights and instead keep the maximum (max pooling) or the mean (average pooling) within a small window (usually $2\times2$ pixels) sliding over the input feature map. Here we use max pooling. 

In Fig.~\ref{Fig: Architecture} we present a schematic overview of the CNN architecture we use for LSBG/artifact classification, that we call \textit{DeepShadows}. It is further described in more detail in Table \ref{table:arch}. \textit{DeepShadows} is a simple sequential architecture, consisted of three convolutional layers (yellow) alternating with pooling layers (red); after the last pooling layer the array is flattened and followed by two fully-connected layer (purple), the last one being a single neuron that outputs the probability (0 to 1) that the input image contains an LSBG.
All convolutional layers use kernels of size $3 \times 3$, while the pooling layers use kernels of size $2 \times 2$. 
Between each convolutional and pooling layer we perform batch normalization \citep{Ioffe2015} to make training faster and more stable. 

To tackle overfitting we employ the following methods: first, we use dropout \citep{Srivastava} after each pooling layer. Dropout sets a specific fraction (here we use 0.4) of randomly selected weights equal to zero. We also use L2  (also known as ridge or Tikhonov) regularization \citep[e.g.,][]{Hastie2020} applied on the weights of the convolutional layers with a penalty term $\lambda = 0.13$ and on the first fully connected layer with penalty $\lambda = 0.12$.
We provide more training details for the \textit{DeepShadows} model in Sec.~\ref{sec: specific_res}.

\section{Classification Results}
\label{sec: Results}

\subsection{Classification Metrics}

To evaluate and compare the performance of the classifiers used in this work we use a number of useful classification metrics, each of which quantifies a different aspect of what a ``good" classification is. For binary probabilistic classifiers, we assume (unless otherwise specified) that an example with output probability $P_{\mbox{\scriptsize{out}}}>0.5$ is classified as an LSBG, while an example with $P_{\mbox{\scriptsize{out}}}<0.5$ is classified as an artifact. We also refer to the LSBGs as the positive class [1] and artifacts as the negative class [0].

True positives (TP) are the correctly classified positive examples, and we analogously define the true negatives (TN), false positives (FP) and false negatives (FN). All the classification metrics, in a binary setting, can be expressed as combinations of these quantities.

The Receiver Operating Characteristic (ROC) curve is a commonly used graphical way to evaluate the performance of a binary classifier; the true positive rate (TPR = TP/(TP+FN)) is plotted versus the false positive rate (FPR = FP/(FP+TN)) at different output probability thresholds (between 0 and 1). A perfect classifier would yield a point in the upper left corner (0,1). A derived quantity is the Area Under the Curve (AUC), with a value closer to 1 signaling a better classifier. ROC curves are useful for visual inspection of the performance of different classifiers and of their uncertainties.

One of the most widely used evaluation metrics is the {\it accuracy}, which measures the fraction of the correct predictions among the total sample examined:
\begin{equation}
    accuracy = \frac{\mbox{TP+TN}}{\mbox{TP+TN+FP+FN}}.
\end{equation}
However, specific problems or applications may have specific requirements that are not fully captured in the overall accuracy. For example, we may be interested in the \textit{completeness} of our classification, in other words, what fraction of the LSBGs were actually classified as such (this metric is also known as \textit{recall} in the machine learning community):
\begin{equation}
completeness = \frac{\mbox{TP}}{\mbox{TP+FN}}.    
\end{equation}
Another useful quantity is the \textit{purity} of the classification: the fraction of objects classified as LSBGs that are true LSBGs (this quantity is also known as \textit{precision}):
\begin{equation}
purity = \frac{\mbox{TP}}{\mbox{TP+FP}}.    
\end{equation}

Finally, we also present the confusion matrix, which includes all four TP, TN, FP, FN values. The confusion matrix can be used to construct a number of other classification metrics.

\subsection{Results}
\label{sec: specific_res}

\begin{figure}[!ht]
\centering
\includegraphics[width=1.0\columnwidth]{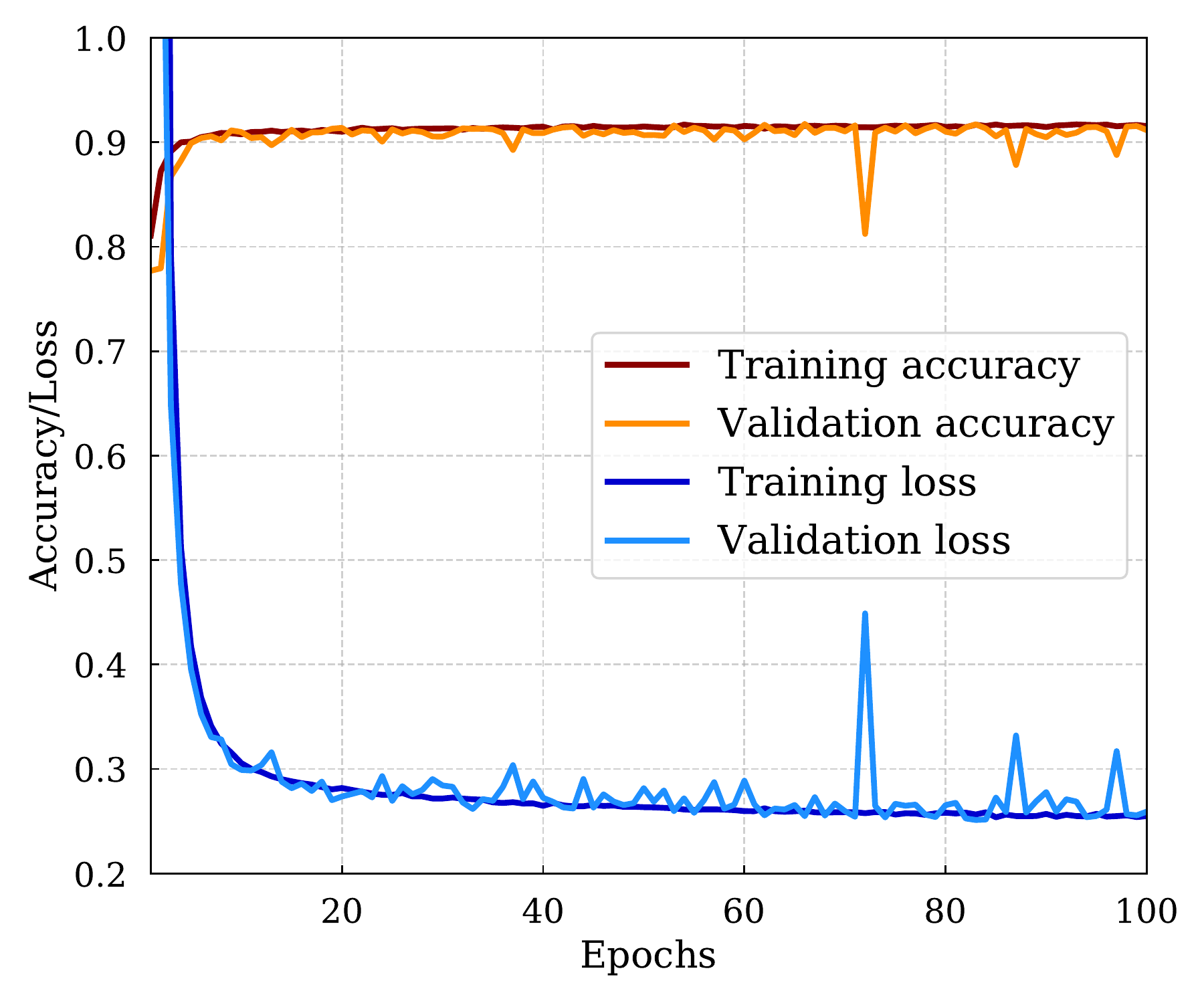}
\caption{Training and validation accuracy/loss as a function of the training epoch for the \textit{DeepShadows} model. Training was performed on 30,000 images and validation on 5,000. The model reaches a training accuracy of $\sim 92\%$ after 100 epochs.}
\label{Fig: History}
\end{figure}

We start by considering the classification results from the two machine learning models (SVMs with an RBF kernel and random forests) described in Sec.~\ref{sec: Machine_Learning}. We use these classification algorithms as implemented in the Python library \texttt{scikit\_learn} \citep{scikit-learn}.\footnote{\url{https://scikit-learn.org/stable/index.html}}

We train these models on the dataset of features derived from \texttt{SourceExtractor}, as described in Sec.~\ref{sec: datasets}. The hyperparameters of the models were tuned by searching a grid of values and using five-fold cross validation on the validation set. The best values were found to be $C=10^4$, $\gamma=0.001$ for the SVM model, while for the random forests model we tune the number of trees in the forest (\texttt{n\_estimators} = 100) and the number of samples required to split internal nodes (\texttt{min\_samples\_split}=10).

The performance of the models was evaluated on the test set. SVMs reach slightly higher accuracy ($81.9\%$ vs $79.7\%$), higher completeness (or recall, $86.7\%$ vs $80.4\%$), similar purity (or precision, $79.6\%$ vs $79.7\%$) and higher AUC (0.894 vs 0.872) compared to the random forest classifier. These results serve as a baseline to compare the performance of our \textit{DeepShadows} CNN model with. 

We implement the \textit{DeepShadows} architecture, as described in Sec.~\ref{sec: Deep_Learning} and Table \ref{table:arch}, using the \texttt{Keras}\footnote{\url{https://keras.io/}} framework on a \texttt{TensorFlow}\footnote{\url{https://www.tensorflow.org/}} backend. We train the model on the training set of 30,000 images described in Sec.~\ref{sec: datasets}. The weights were updated using Adadelta \citep{Adadelta}, an optimized version of the vanilla stochastic gradient descent algorithm, with a learning rate of $\eta=0.1$. The update is performed iteratively in batches of images; we use a batch size of 64. A training \textit{epoch} occurs once every image in the training set has been used to update the network weights. 
We train our model for 100 epochs; we do not continue for more epochs since the results do not improve more. We validate the process on the validation set of 5,000 images described in Sec.~\ref{sec: datasets}. In Fig.~\ref{Fig: History} we present the training history of our model (accuracy/loss as a function of training epoch). We can see that our model converges well and that there are no signs of overfitting or underfitting (the training and validation curves closely follow each other).

\begin{figure*}[h]
\centering
\subfigure[]{\includegraphics[width=0.43\textwidth]{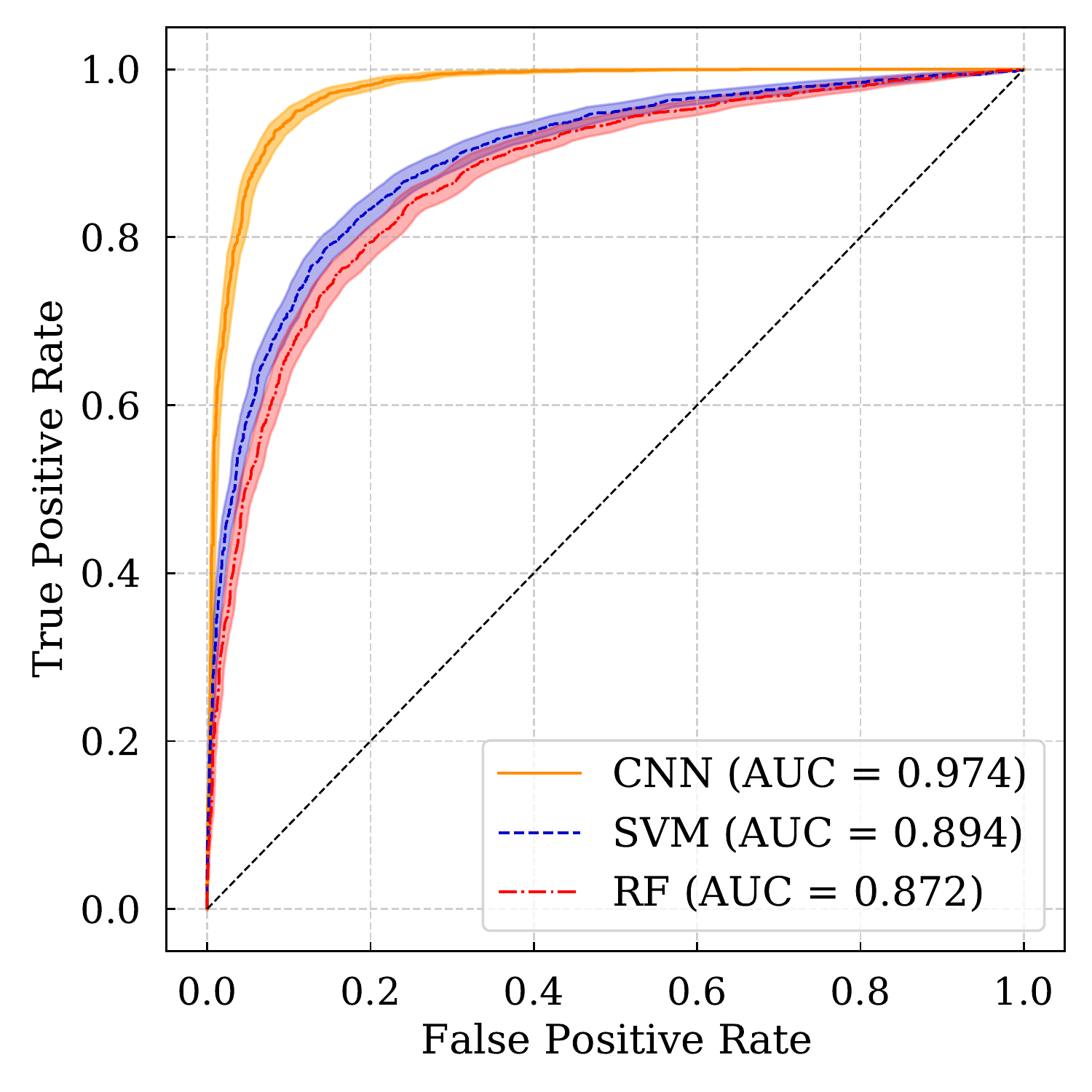}}
\hspace*{\fill}
\subfigure[]{\includegraphics[width=0.52\textwidth]{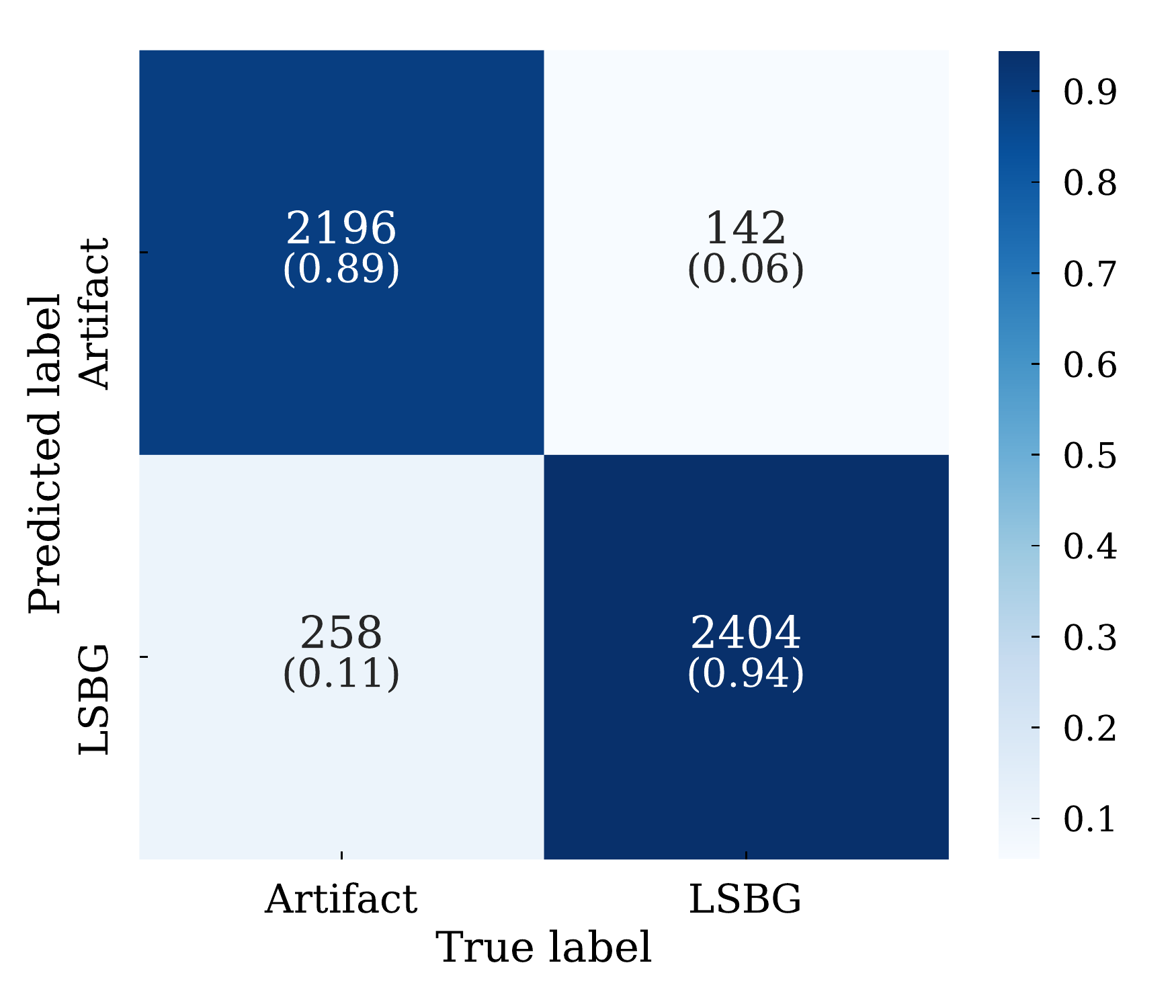}}
\caption{(a) ROC curves for the \textit{DeepShadows} CNN model (orange, solid line), the SVM model (blue dashed line), and the random forest model (red dashed-dotted line) evaluated on the test set. The diagonal dashed line corresponds to the performance of a random classifier. We also show the $95\%$ confidence intervals on the vertical direction (true positive rate). These were estimated using the bootstrap method on the test set of images (see Sec.~\ref{sec: Errors}). (b) Confusion matrix of the \textit{DeepShadows} model predictions on the test set. The values in parentheses correspond to the normalized version of the matrix, obtained by dividing the raw number of objects in each case by the total number of objects in each category (true label).
\label{fig: ROC_and_Confusion}
}
\end{figure*}

\begin{table}[!ht]
\caption{Comparison of the classification metrics for the three machine learning models presented in Sec. \ref{sec: specific_res}.}
\label{table: perf_comparison}
\centering
\begin{tabular}{|l||c|c|c|}
\hline
\diaghead{\theadfont Diag ColumnmnHead I}%
{\textbf{{\normalsize{Metric}}}}{{\normalsize{\textbf{Model}}}}& \textbf{SVM}& \textbf{RF} & \textbf{CNN} \\
\hline \hline
Accuracy & 0.819& 0.797 & 0.920\\
Completeness & 0.867& 0.804 & 0.944\\
Purity & 0.796& 0.797 & 0.903\\
AUC score & 0.894& 0.872 & 0.974\\
\hline
\end{tabular}
\end{table}

The \textit{DeepShadows} CNN classifier reaches an accuracy of $92.0 \%$, completeness (recall) of $94.4 \%$ and purity (precision) of $90.3 \%$ and AUC score equal to $0.974$, all evaluated on the test set of $5,000$ images. These values are significantly higher than those obtained from the SVM and random forest models (see also Table \ref{table: perf_comparison} for a direct comparison). Note that these classical machine learning models were trained on a physically motivated set of features that is not guaranteed to be optimal, while \textit{DeepShadows} works directly at the pixel level.

The fact that \textit{DeepShadows} is a more powerful classifier can be visually demonstrated by plotting the ROC curves of the three models (left panel of Fig.~\ref{fig: ROC_and_Confusion}). In the same figure we also show the AUC scores, as well as the $95\%$ confidence intervals on the ROC curves, estimated using the bootstrap method on the test set (see Sec.~\ref{sec: Errors}).

On the right-hand side of Fig.~\ref{fig: ROC_and_Confusion} we present the confusion matrix for \textit{DeepShadows} that shows the number of the correctly classified and misclassified objects. Many common classification metrics can be derived from the confusion matrix. In parentheses we present the entries of the normalized confusion matrix, which can be obtained by dividing by the total number of objects in each (true label) category. We see that most misclassification cases occur in artifacts classified as LSBGs, something that is also evident from the lower value of purity compared to completeness.

\subsection{Interpretation of Results}
\label{sec: Interpretation}

\begin{figure}[!ht]
\centering
\includegraphics[width=0.9\columnwidth]{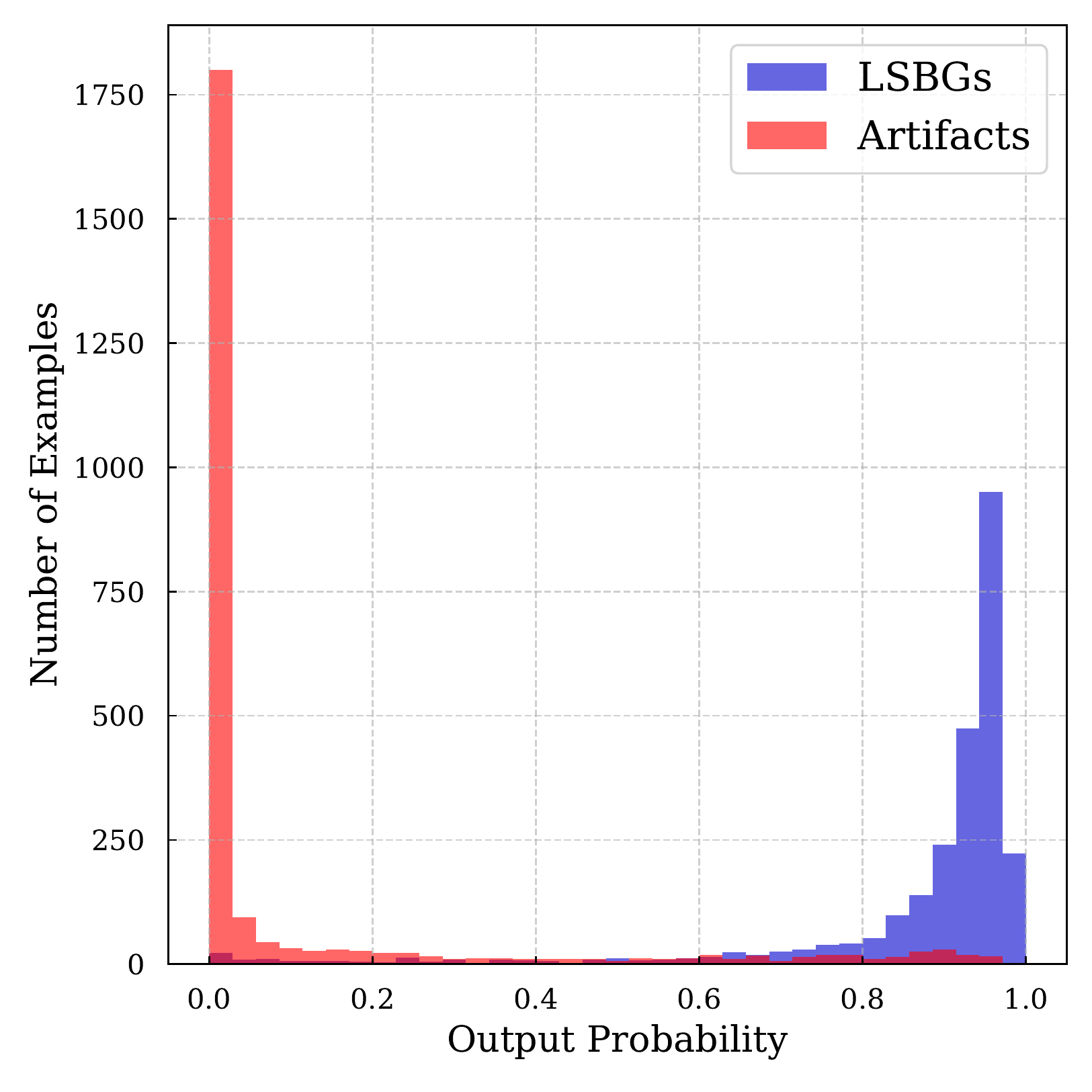}
\caption{Output probabilities from the \textit{DeepShadows} CNN for the LSBGs and artifacts in the test set.}
\label{Fig: Probas}
\end{figure}

\begin{figure*}[!ht]
\centering
\subfigure{\includegraphics[width=0.243\textwidth]{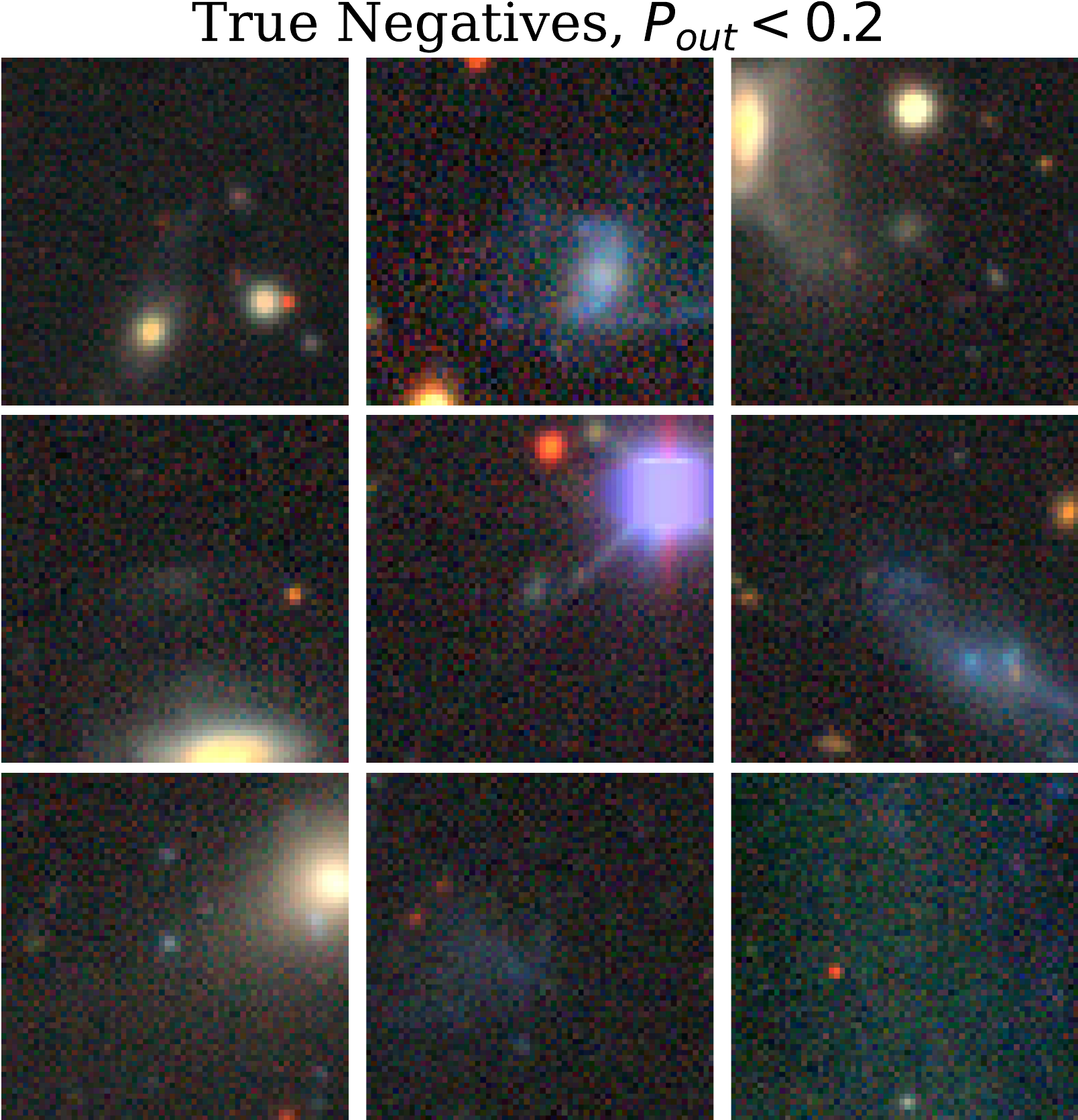}}
\subfigure{\includegraphics[width=0.243\textwidth]{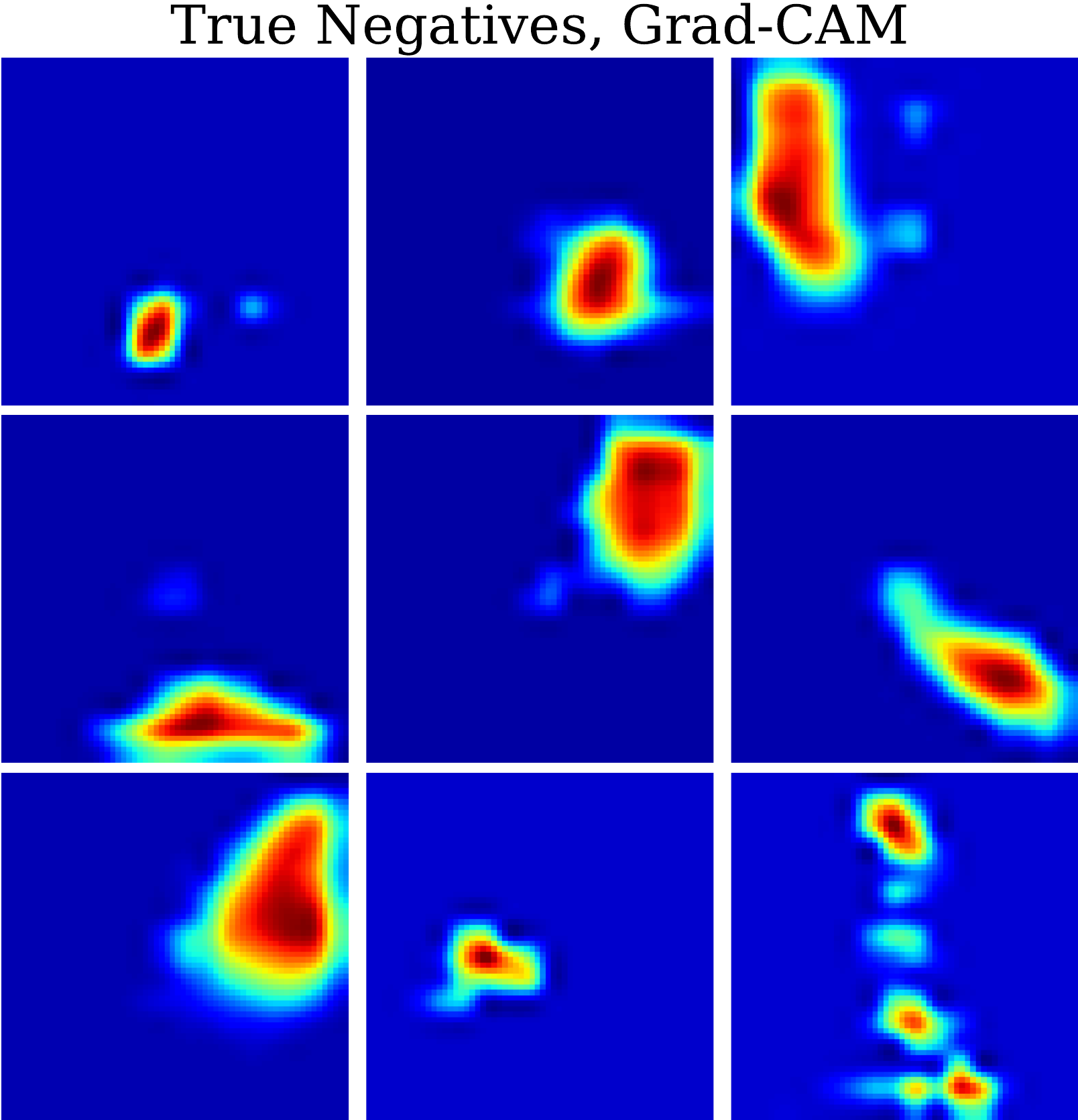}}
\hspace*{\fill}
\subfigure{\includegraphics[width=0.243\textwidth]{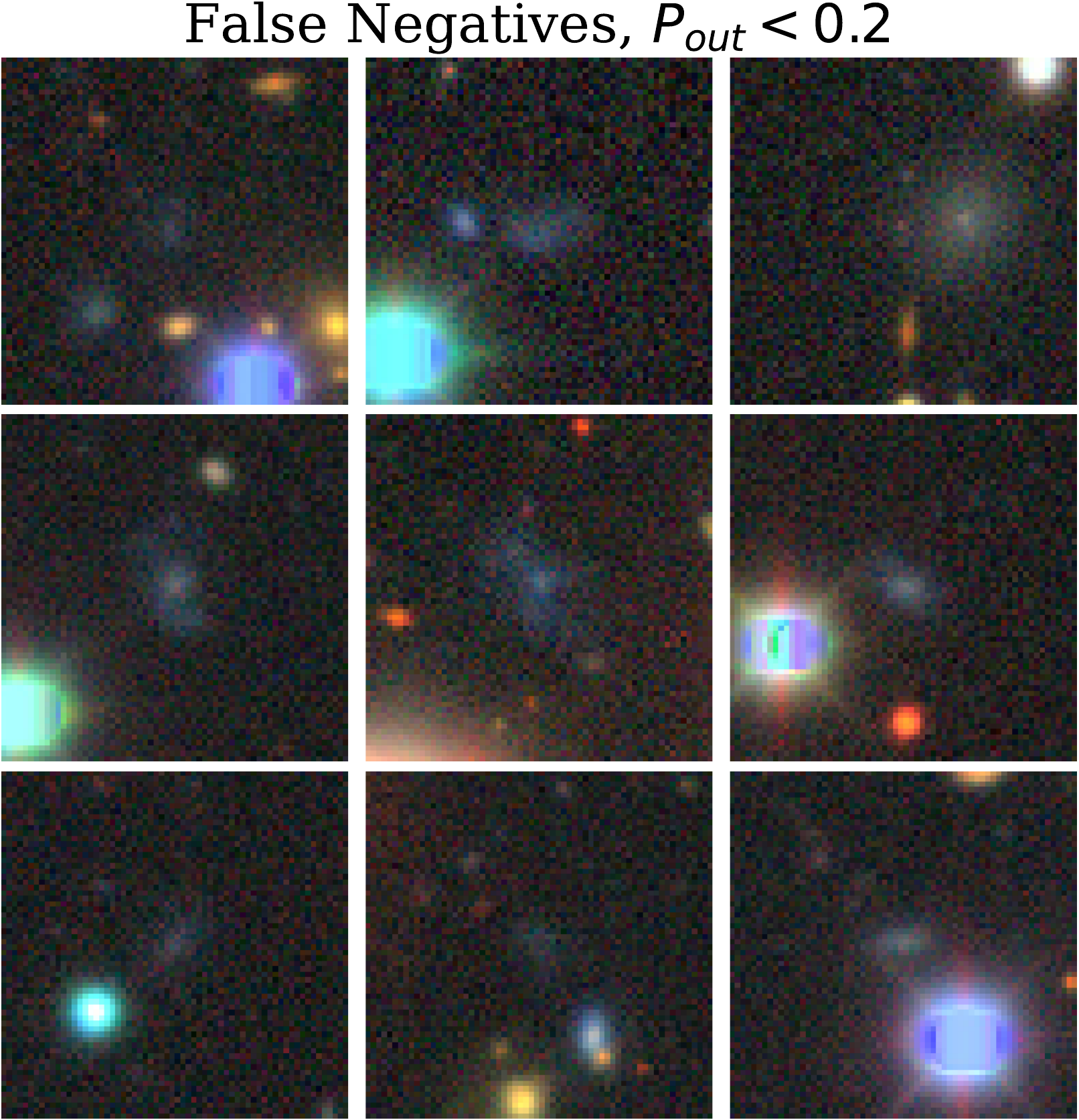}}
\subfigure{\includegraphics[width=0.243\textwidth]{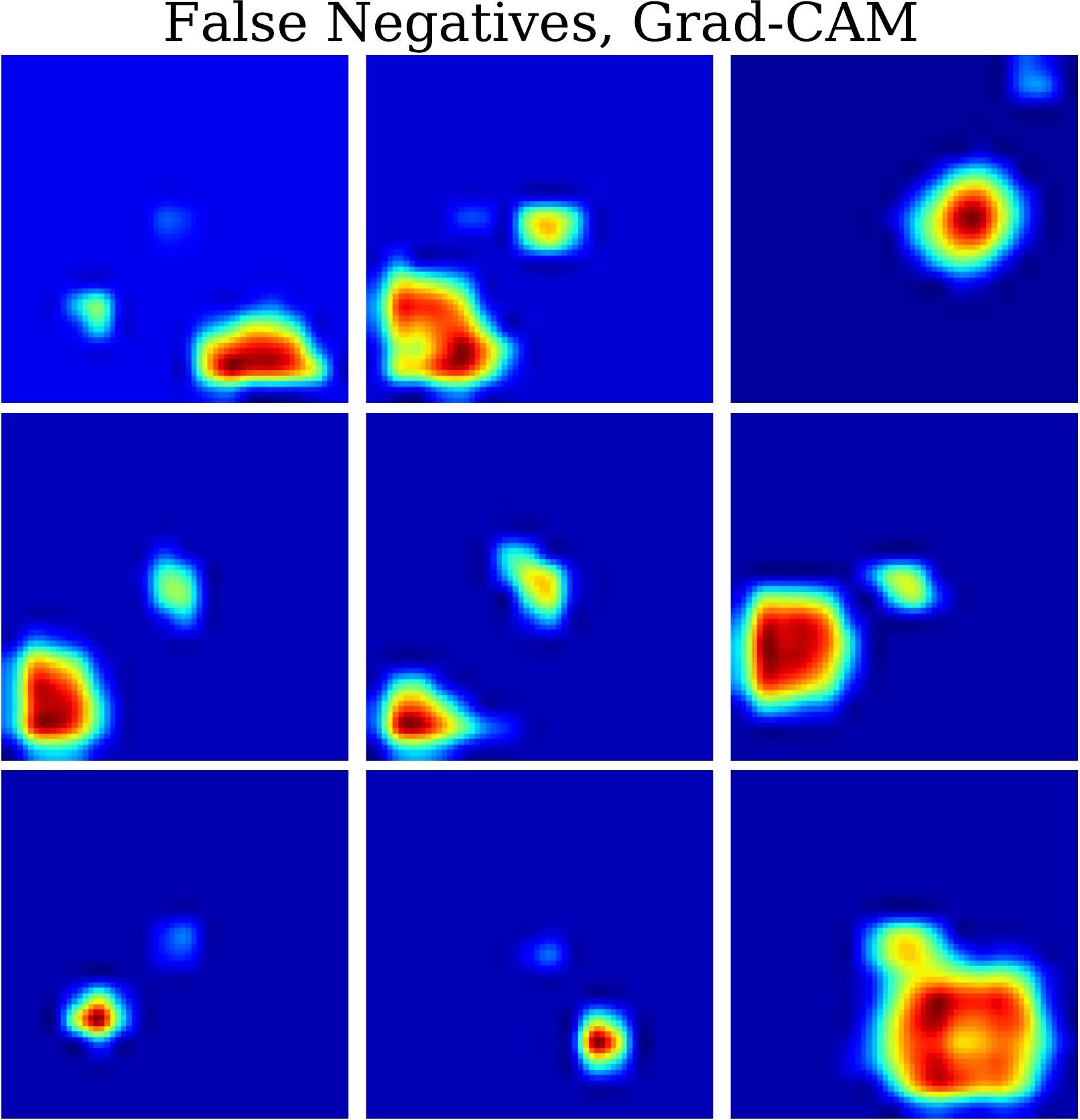}}\\
\vspace{0.25cm}
\subfigure{\includegraphics[width=0.243\textwidth]{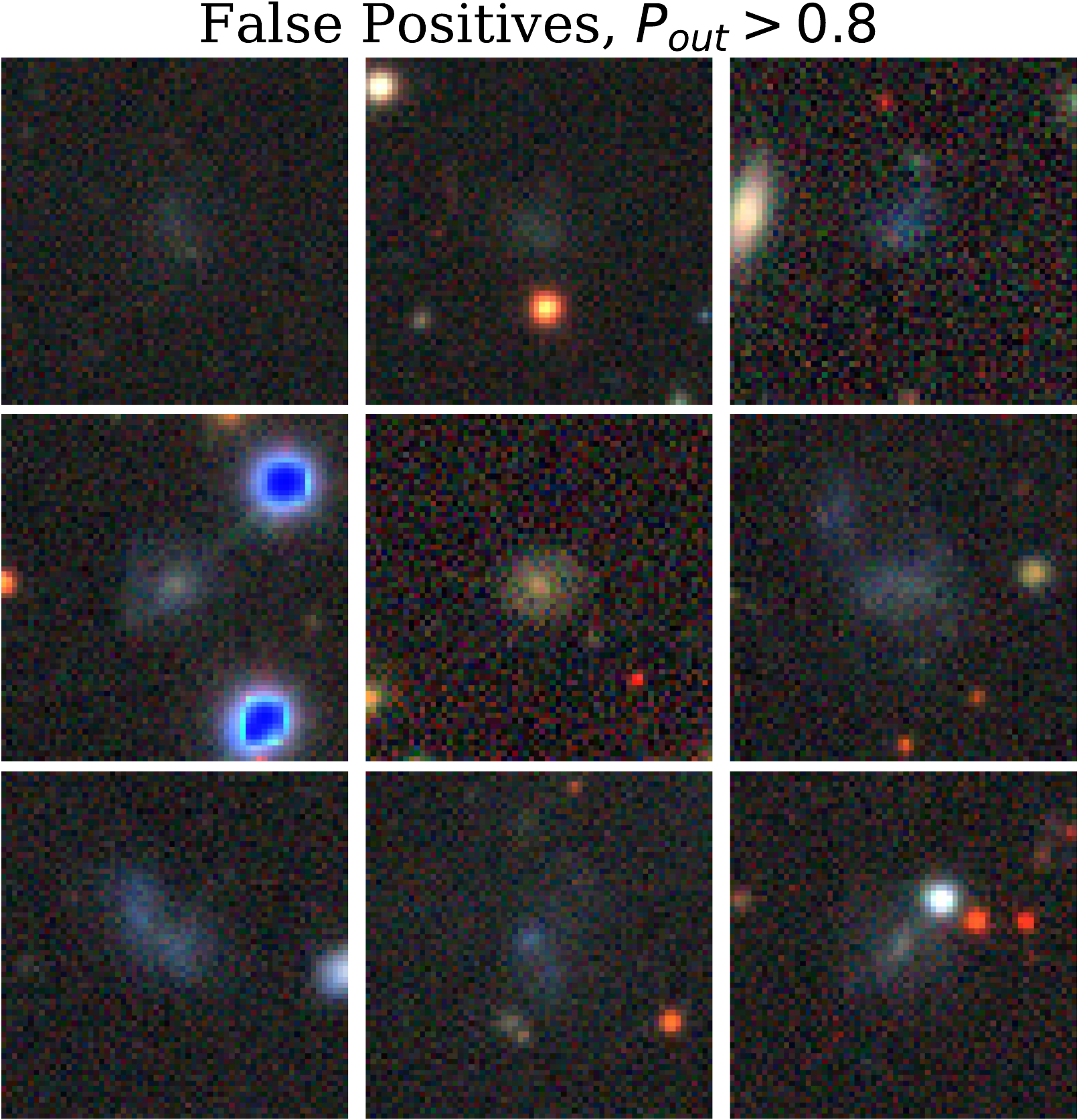}}
\subfigure{\includegraphics[width=0.243\textwidth]{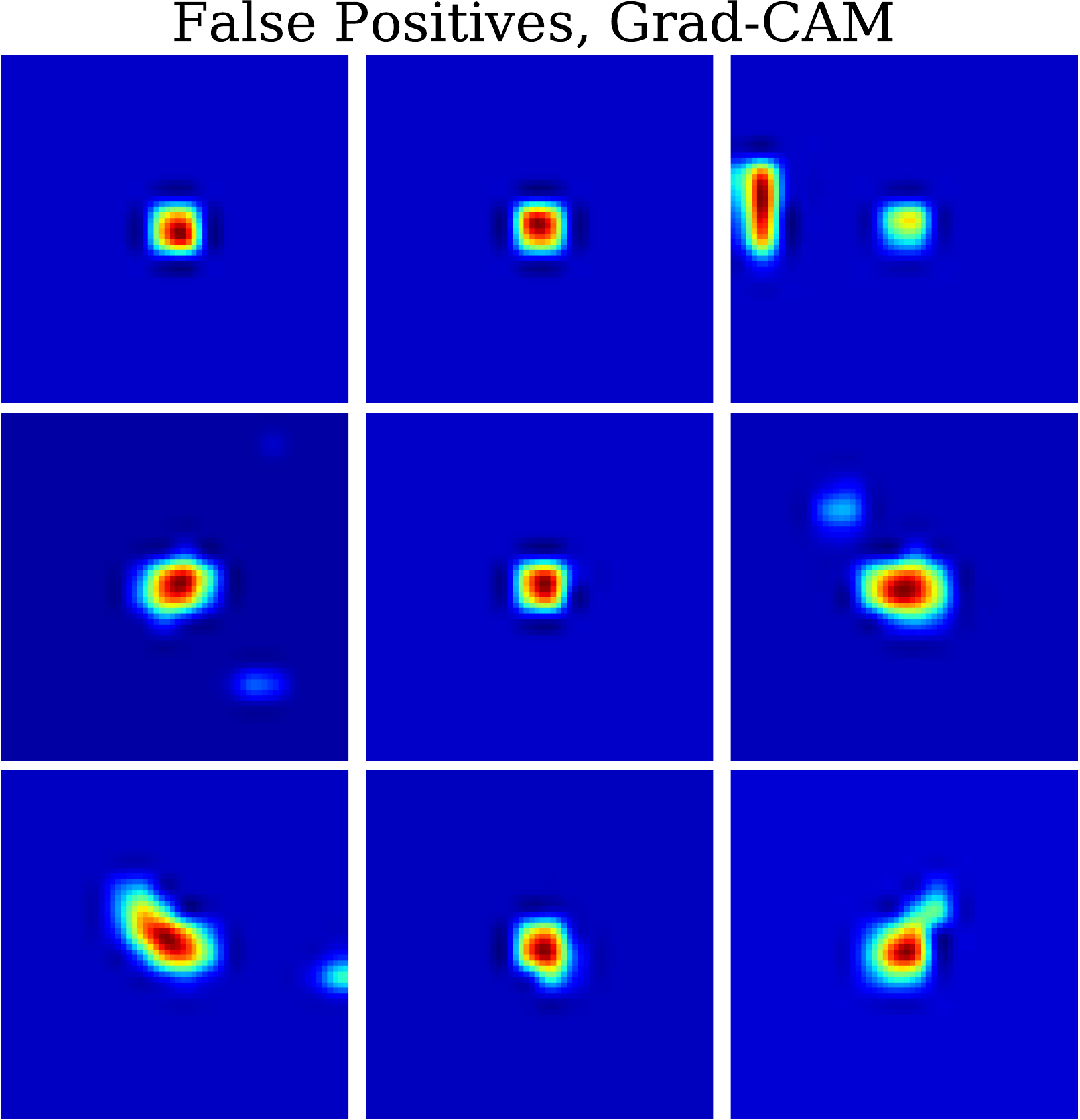}}
\hspace*{\fill}
\subfigure{\includegraphics[width=0.243\textwidth]{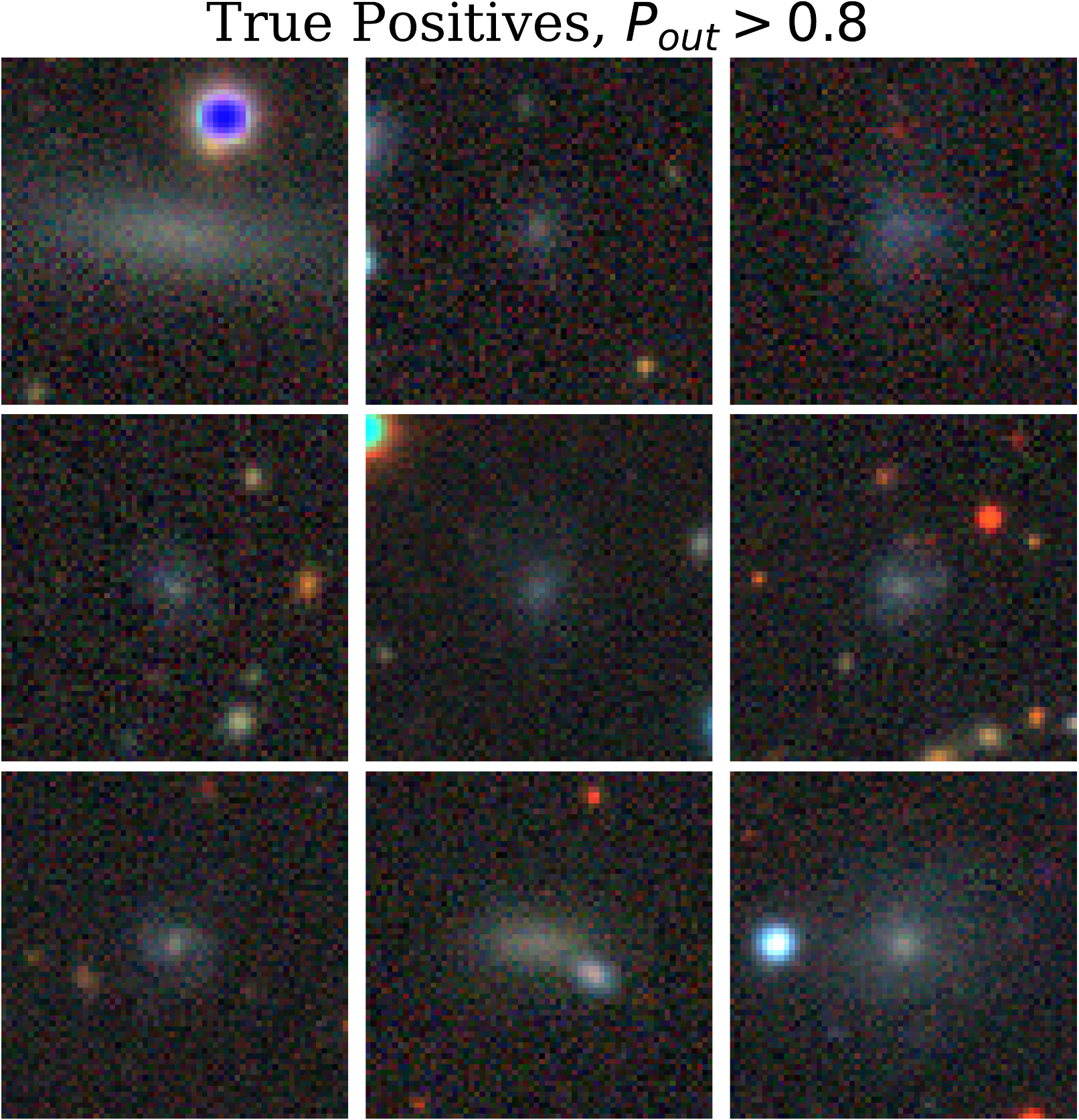}}
\subfigure{\includegraphics[width=0.243\textwidth]{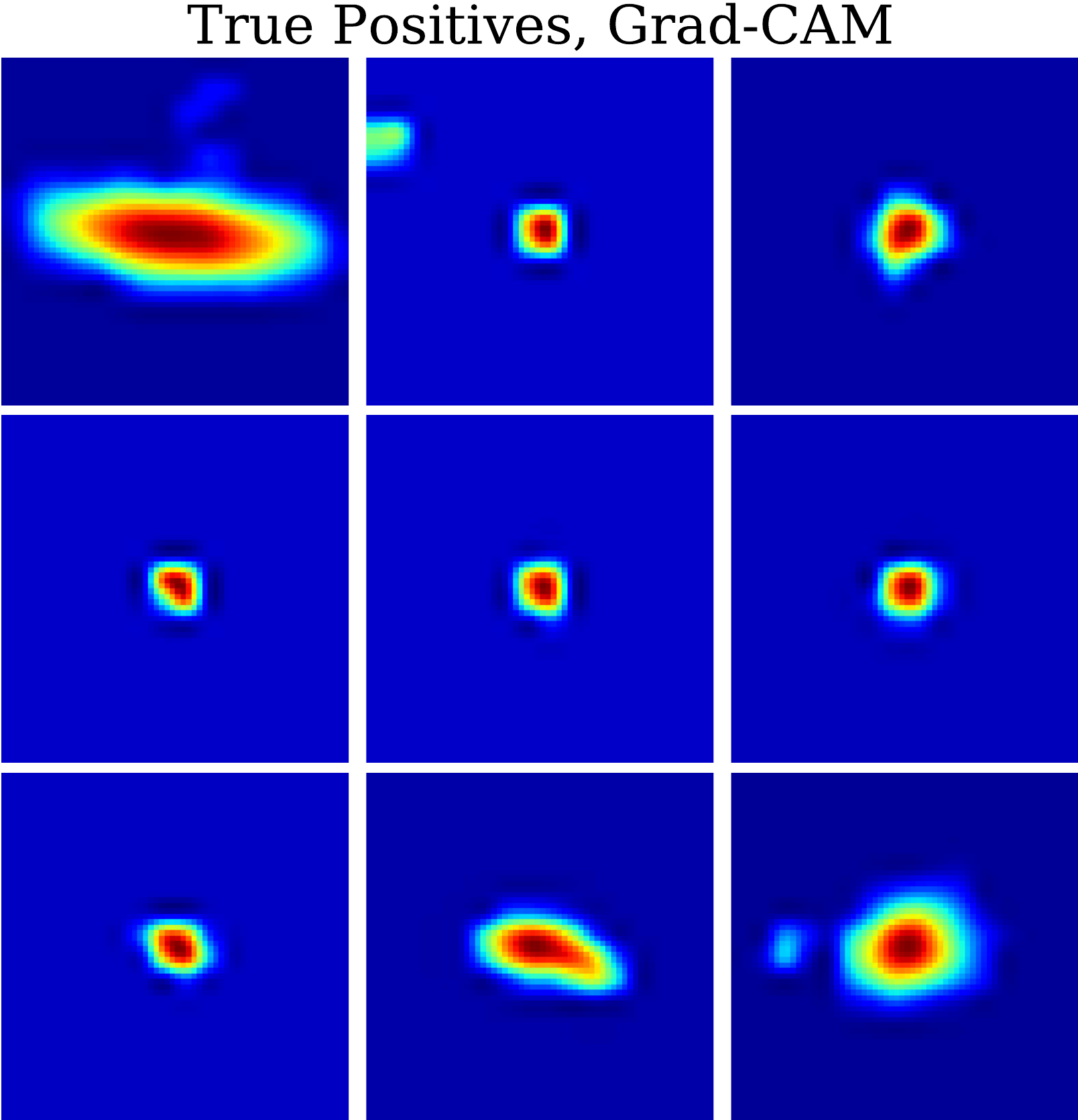}}
\caption{Examples of objects classified with high probability in their respective class and corresponding Grad-CAM visualization maps for the same objects. Clockwise: True Negatives, False Negatives, True Positives, False Positives (same arrangement as the confusion matrix).}
\label{fig: Grad_CAM}
\end{figure*}

The \textit{DeepShadows} CNN is a probabilistic classifier that outputs the probability that an example is an LSBG. The classification results presented so far assume a probability threshold  $P_{\mbox{\scriptsize{out}}}=0.5$ (any image with higher output probability is classified as an LSBG, while any image with lower output probability is classified an artifact). 
We can get more insight about the classification outcomes, and try to interpret the results, by examining the predicted probabilities for the objects in the test set.
We plot these probabilities in Fig.~\ref{Fig: Probas}. 
Output probabilities of those objects with true label ``artifact" are in red and those with true label ``LSBG" are in blue. 
Artifacts are found to be more concentrated towards the $P_{\mbox{\scriptsize{out}}}=0$, implying that most can be easily distinguished from LSBGs. 
However, there is also a long tail in this distribution with some objects that are labeled (true value) as artifacts but have been assigned a high probability of being LSBGs.
LSBGs, on the other hand, have a wider  distribution in probabilities (less concentrated towards $P_{\mbox{\scriptsize{out}}}=1$) but a less significant tail to very low probabilities.

To better understand our results it is useful to inspect some of the objects that: (a) were assigned to the wrong class but with high confidence, or (b) were assigned to the  correct class with high confidence. 
To interpret the classification results we employ a recent technique, called Gradient-weighted Class Activation Mapping \citep[Grad-CAM,][]{GRADCAM}. Grad-CAM allows us to produce ``visual explanations" for the classification results, by highlighting the most important regions in the classification procedure.
We provide more technical details about Grad-CAM in \ref{sec: Grad_CAM}; here we just note that these images are produced by calculating the gradients of the feature maps of the last convolutional layer with respect to the output score for each  class. 

In Fig.~\ref{fig: Grad_CAM} we present examples and corresponding Grad-CAMs for randomly selected (clockwise): true negatives, false negatives, true positives, false positives. All these examples were classified with high confidence to their assigned categories ($P_{\mbox{\scriptsize{out}}}>0.8$ for those classified as LSBGs and $P_{\mbox{\scriptsize{out}}}<0.2$ for those classified as artifacts).

The images of the objects classified as negatives (artifacts), both true and false, are characterized by the presence of off-centered light sources (such as stars or components of galaxies). The fact that these are the important regions for the classification problems is also confirmed by Grad-CAM maps. Especially in the case of false negatives, we see that the central LSBGs are shadowed by the presence of other nearby objects that contribute to the decision of \textit{DeepShadows} to classify them as artifacts. 

On the other hand, the images of those objects classified as LSBGs (positive class) are dominated by the presence of a central object; this can also be seen in the highlighted regions of the Grad-CAM maps. Interestingly, the high-confidence false positives presented here seem to be real galaxies. These objects were likely rejected out of an abundance of caution when visually selecting LSBGs, since these objects were generally more compact or faint compared to other LSBGs. The neural network classifier is able to ``correct" the human labeling.

\section{Transfer Learning}
\label{sec: Transfer_learning}

\begin{figure*}[!ht]
\centering
\subfigure[]{\includegraphics[width=0.40\textwidth]{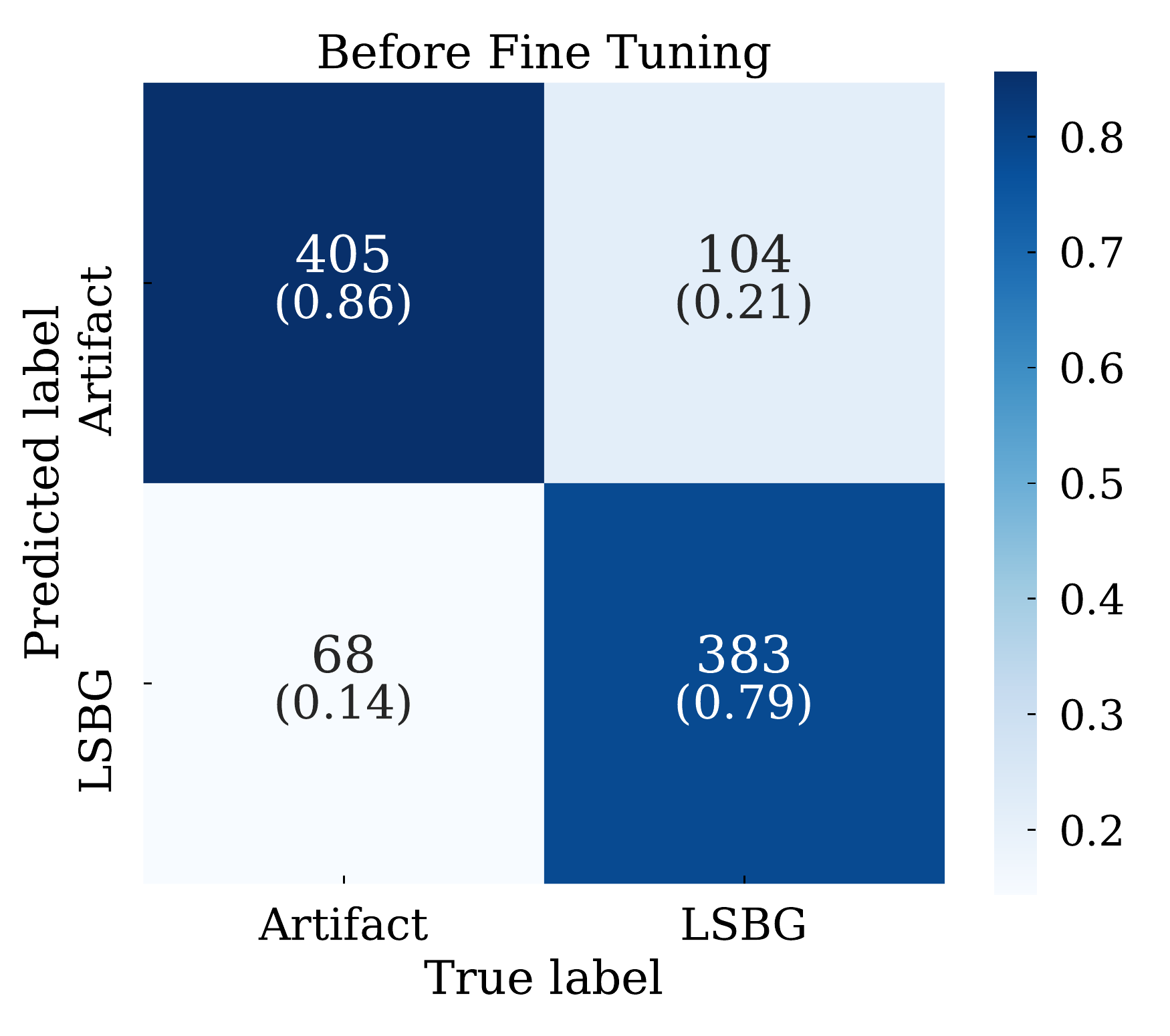}}
\subfigure[]{\includegraphics[width=0.40\textwidth]{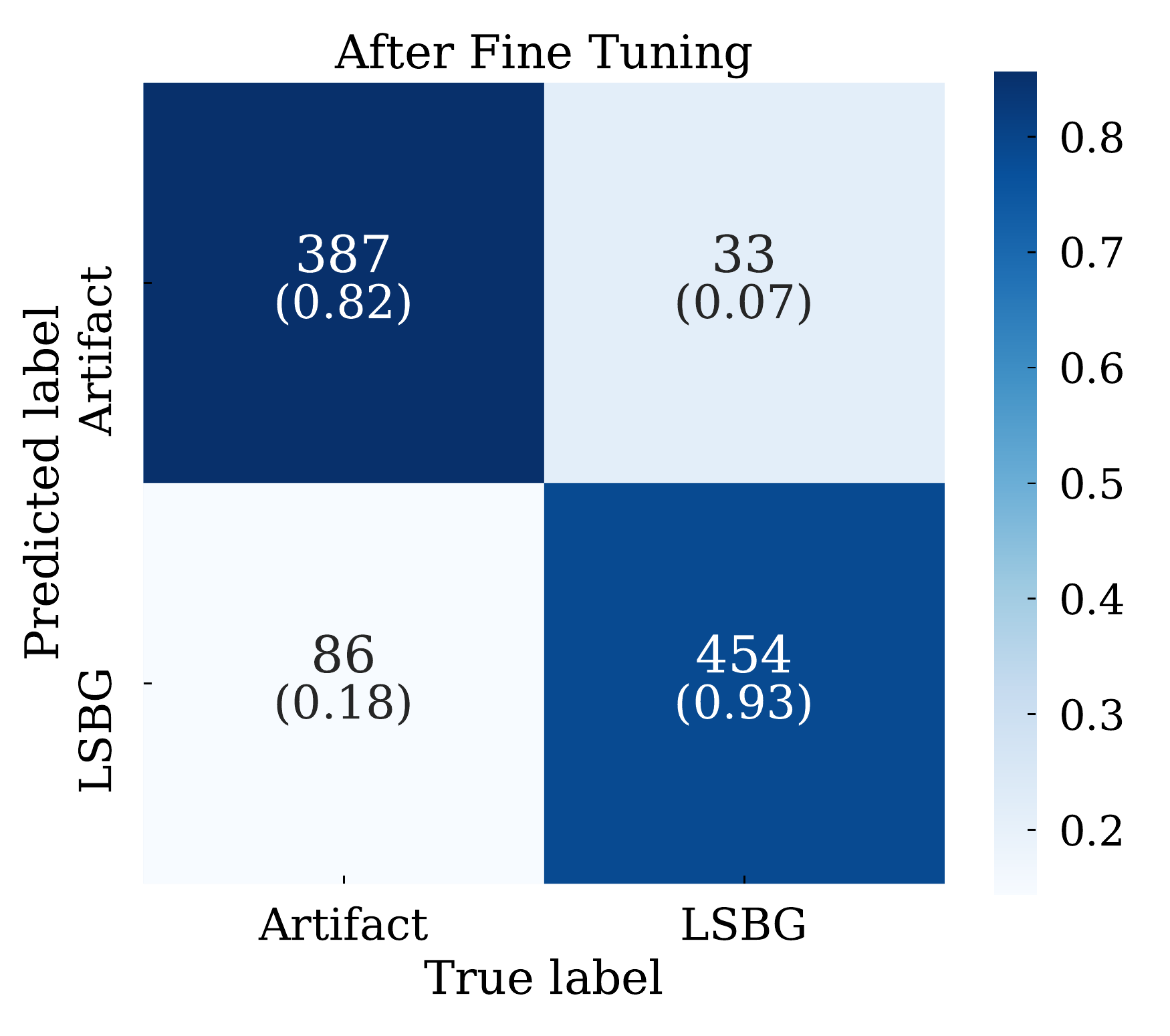}}\\
\vspace{0.2cm}
\subfigure[]{\includegraphics[width=0.43\textwidth]{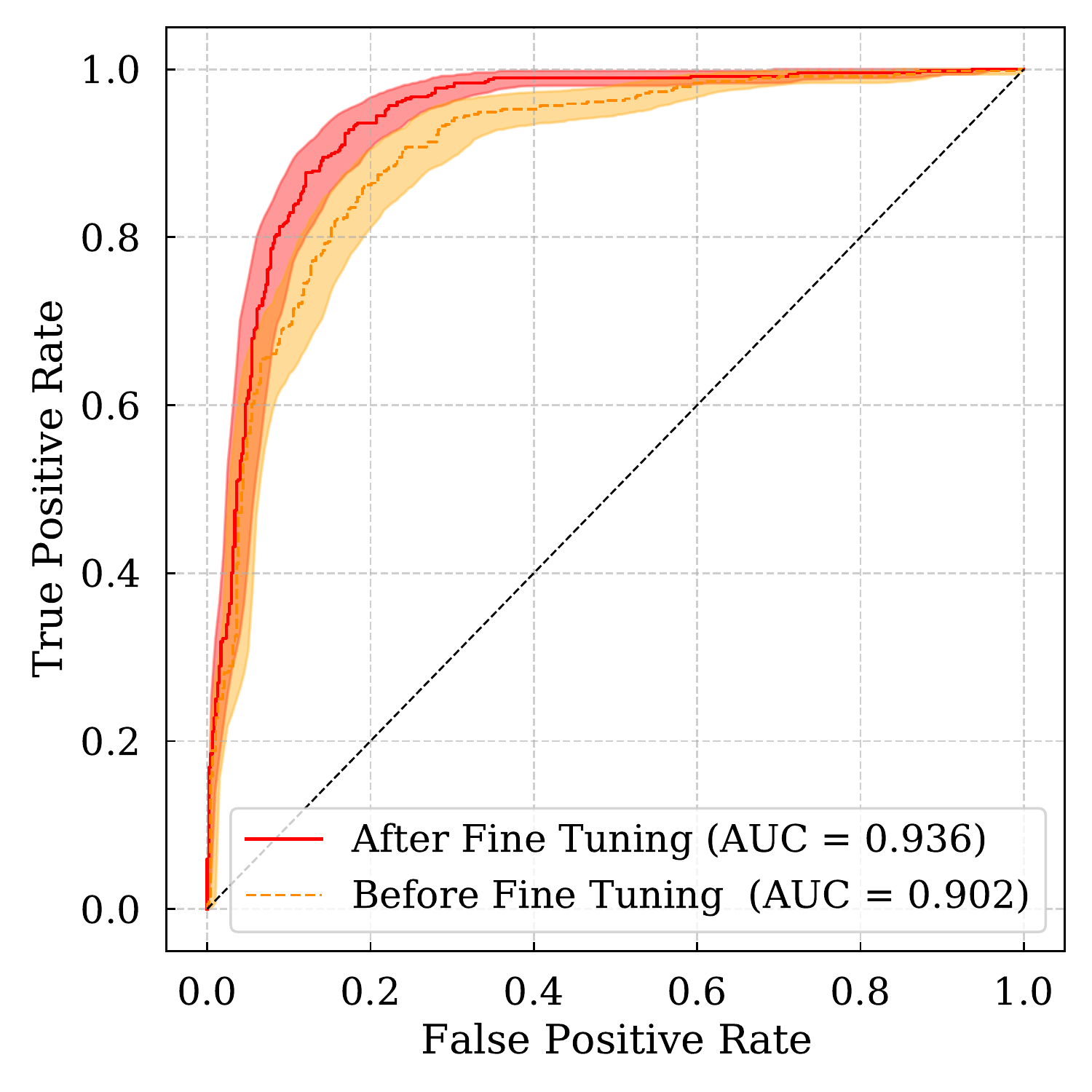}}
\vspace{-0.1cm}
\caption{Transfer learning results on the HSC SSP test set. (a) Confusion matrix (raw and normalized values) before fine tuning. (b) Confusion matrix (raw and normalized values) after fine tuning. (c) ROC curves and AUC scores, before (orange, dashed line) and after (red, solid line) fine tuning. The bands correspond to the $95\%$ confidence level intervals.}
\label{fig: HSC_results}
\end{figure*}

In the previous section, both the training and evaluation of the classifiers used data from the same survey, namely DES. 
These results demonstrate that a CNN classifier trained on a large training sample can be used to separate LSBGs from artifacts with a high accuracy (and purity/completeness). However, a \textit{large} number of labeled training examples is required. In this section we explore whether we can a classifier trained on one survey and apply it to data from another survey. If such an approach is successful, it can significantly reduce the need to generate large training sets  via visual inspection in future surveys. 

\textit{Transfer learning} refers to the process of training a machine learning algorithm to perform a task and then using it to perform another related task or perform the same task on a dataset with different specifications \citep[e.g.,][]{Weiss2016}. Transfer learning has found many applications, including image recognition problems \citep[e.g.,][]{Pan_Yang2010,Bengio2012,Yosinski2014,Zhuang2019}. Its power has recently been explored in astronomy \citep{Vilalta2018}, especially in the context of cross-survey classification, namely in the field of galaxy morphology prediction \citep{Domingues_Sanchez2019}. 
The use of transfer learning has also been investigated for classification of galaxy mergers \citep{Ackermann2018}, radio galaxy classification \citep{Tang2019}, star-galaxy classification \citep{Wei2020}, even glitch classification of LIGO events \citep{Gliches}.

Here we use data from the HSC SSP, for which we have a small visually classified sample of LSBGs (Sec.~\ref{sec: HSC}), to study how successfully a model trained on one survey can be used to distinguish between LSBGs and artifacts in another survey.
Following \citet{Domingues_Sanchez2019} we consider two cases:
\begin{enumerate}[a)]
    \item Apply the \textit{DeepShadows} model, which was trained on DES data, directly to the HSC SSP data (test set) without any further training.
    \item Before predicting on the HSC SSP test set, we use a small set of 320 objects from the HSC SSP to perform a \textit{fine-tuning} step. We re-train the whole model using a much smaller learning rate (in order to keep the change of the weights low and avoid overfitting), $\eta= 0.005$, for 30 epochs, using a batch size of 16 (Note that, alternatively, one can re-train only the final, dense layers. We do not consider this case here).
\end{enumerate}

We present the results (confusion matrices and ROC curves) for the two cases in Fig.~ \ref{fig: HSC_results}. Before fine tuning, \textit{DeepShadows} has an accuracy of $82.1\%$, purity (precision) of $84.9\%$ and completeness (recall) of $78.6\%$. Classification performance significantly improves with fine tuning, as can be seen by inspecting the two ROC curves (and the corresponding AUC scores) in panel (c) of Fig.~\ref{fig: HSC_results}. The better performance is driven by the increased number of true positives and correspondingly smaller number of false negatives, as can be seen by comparing the confusion matrices in panels (a) and (b) of Fig.~\ref{fig: HSC_results}. The overall accuracy after fine tuning reaches a value of $87.6\%$, the purity a value of $84.1\%$ and the completeness an impressive $93.2\%$ (almost as good as the application to the one we get from applying to the DES data).

\section{Uncertainty Quantification}
\label{sec: Errors}

\begin{figure*}[h]
\centering
\subfigure[]{\includegraphics[width=0.47\textwidth]{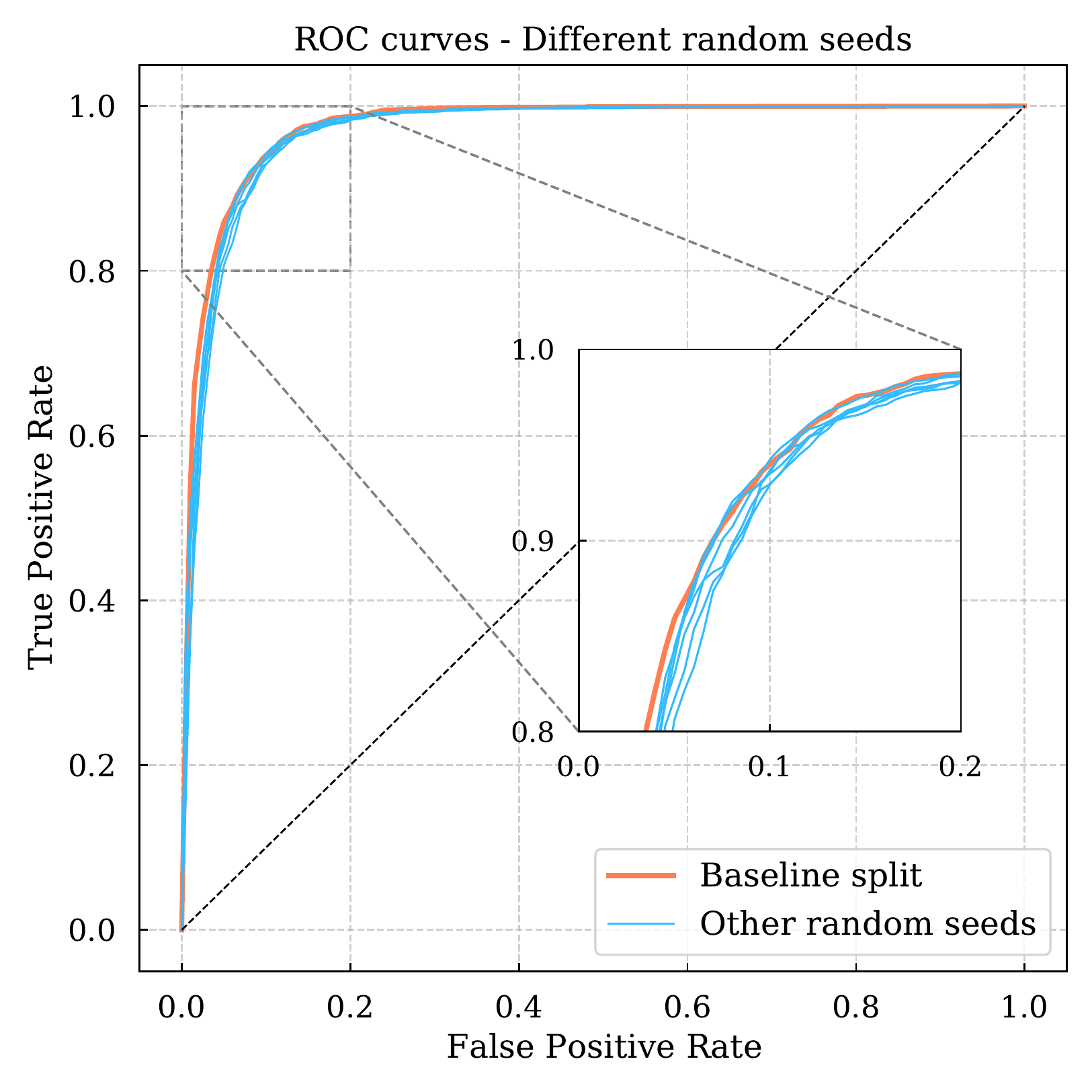}}
\hspace*{\fill}
\subfigure[]{\includegraphics[width=0.47\textwidth]{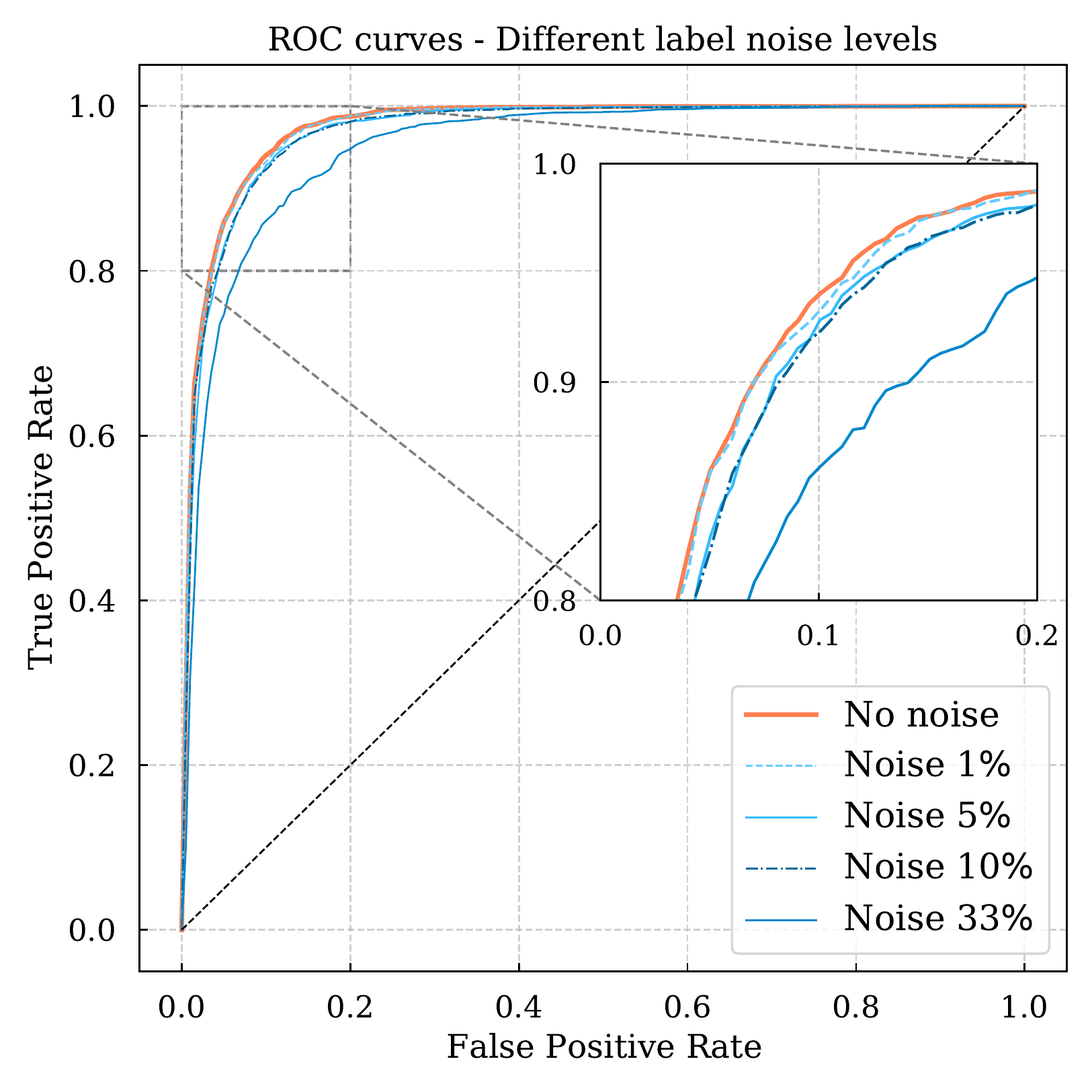}}
\caption{(a) ROC curves of the performance on the test set using the baseline split (orange line) between training-validation-test sets and using splits with different random seeds (blue lines). (a) ROC curves of the performance on the test set using different levels of label noise in the training set, starting from no noise (orange line) to $33\%$ random mislabeling.}
\label{fig: Uncertainties}
\end{figure*}

We have presented performance metrics for our models without discussing potential uncertainties in these estimates. 
We now consider three potential sources of uncertainty:
\begin{enumerate}[a)]
    \item Random statistical uncertainties when calculating the evaluation metrics on the test set (subsection \ref{sec: bootstrap}).
    \item Uncertainties arising from randomly splitting the data into training, validation, and test sets (subsection 
    \ref{sec: split}).
    \item Label noise from the presence of mislabeled examples in our training set (subsection \ref{sec: mislabeling}).
\end{enumerate}

Our analysis is not intended to be exhaustive. For a more complete discussion of the uncertainties present in deep learning models see e.g., \citet{Kendall2017}, \citet{Hullermeier2019}, \citet{Caldeira_Nord2020}, and references therein.

\subsection{Bootstrap resampling of test set}
\label{sec: bootstrap}

We estimate $95\%$ confidence intervals on the classification metrics by bootstrapping  the test set 1000 times with replacement and classifying each realization. 
We present these confidence intervals in Table \ref{table:Bootstap}; the $95\%$ intervals for the true positive rate are also presented in panel (a) of Fig.~\ref{fig: ROC_and_Confusion}.
Note that this approach is not comprehensive. 
In principle, we could have bootstrapped the training set to evaluate model error in addition to statistical prediction error. 
However, re-training \textit{DeepShadows} 1000 times was computationally prohibitive.

\begin{table}[!ht]
\caption{95$\%$ Confidence intervals on the classification metrics from Bootstrap resampling of the test set.}  
\label{table:Bootstap}
\centering
\begin{tabular}{|l | c |}
\hline
\textbf{Metric} & $\mathbf{95 \%}$ \textbf{Confidence Interval}\\
\hline
Accuracy & 0.912 -- 0.928 \\
Completeness & 0.935 -- 0.953 \\
Purity & 0.891 -- 0.914 \\
AUC score & 0.970 -- 0.974\\
\hline
\end{tabular}
\end{table}

The typical interval for all parameters is $({\sim} 1.5\%)$. This is much smaller than the difference between the performance of \textit{DeepShadows} and the other machine learning models, robustly demonstrating that \textit{DeepShadows} performs better in classifying galaxies and artifacts. The confidence intervals of the SVM and random forest models are of similar width.

\begin{table}[!ht]
\caption{95$\%$ Confidence Intervals on the classification metrics from Bootstrap for the transfer learning task, after fine-tuning, evaluated on the HSC SSP test set.}  
\label{table:Bootstap_HSC}
\centering
\begin{tabular}{|l | c |}
\hline
\textbf{Metric} & $\mathbf{95 \%}$ \textbf{Confidence Interval}\\
\hline
Accuracy & 0.856 -- 0.896 \\
Completeness & 0.910 -- 0.954 \\
Purity & 0.808 -- 0.871 \\
AUC score & 0.921 -- 0.951\\
\hline
\end{tabular}
\end{table}

We have seen in Fig.~\ref{fig: HSC_results} that the confidence intervals in the ROC curves (true positive rates) in the case where we explored transfer learning are wider than those of the main section (training and test on the DES data). In Table \ref{table:Bootstap_HSC} we present 95$\%$ confidence intervals for the transfer learning task, after fine tuning, evaluated on the HSC SSP test set. As we can see the typical range in the evaluation metrics in this case is $\sim$0.04--0.05, significantly larger than before.

\subsection{DES dataset assignment}
\label{sec: split}

\begin{table*}[!ht]
\caption{Classification metrics for the baseline split of training-validation-test sets and six alternative splits using different random seeds. Because of the small number of runs we present each case individually and not as an interval like in Table \ref{table:Bootstap}.}
\label{table:random_seeds}
\centering
\begin{tabular}{|c||c|c|c|c|}
\hline
\diaghead{\theadfont Diag ColumnmnHead I}%
{\textbf{{\normalsize{Seed $\#$}}}}{{\normalsize{\textbf{Metric}}}}& Accuracy& Completeness & Purity & AUC score \\
\hline \hline
\textbf{Baseline} & 0.920 & 0.944 & 0.903 & 0.974 \\
1st run & 0.912& 0.974 & 0.868 & 0.967\\
2nd run & 0.915& 0.972 & 0.870 & 0.971\\
3rd run & 0.917& 0.950 & 0.889 & 0.968\\
4th run & 0.914& 0.936 & 0.896 & 0.968\\
5th run & 0.920& 0.917 & 0.924 & 0.972\\
6th run & 0.915& 0.937 & 0.898 & 0.971\\
\hline
\end{tabular}
\end{table*}

We have noticed that the model performance is sensitive (at a level similar to the uncertainties described in the previous section) to the random assignment of examples into the training-validation-test sets. 
We study variations in model performance stemming from random assignment by splitting the whole dataset into training-validation-test using 6 different random seeds, retraining using each new training set, and evaluating on the new test sets. Note that we do not change the sizes of these sets; these are always 30,000 (training), 5,000 (validation), 5,000 (test). 

In Table \ref{table:random_seeds} we present the classification metrics for each one of the runs while in the left-hand side of Fig.~\ref{fig: Uncertainties} we show the ROC curves for each one of these cases. Due to the limited number of runs (constrained by the cost of re-training the model each time) we don't present summary statistics (like 95$\%$ confidence intervals);  however we can see that for each metric, the values span a range similar to those presented in Table \ref{table:Bootstap}.
The baseline split, used in Sec.~\ref{sec: Results} was selected as one of the better performing models after several trials. In the GitHub page of this project we make available the on-sky coordinates (right ascension, declination) of the training, validation and test sets under the baseline split.

\subsection{Impact of mislabeling}
\label{sec: mislabeling}

\begin{figure}[!ht]
\centering
\includegraphics[width=0.9\columnwidth]{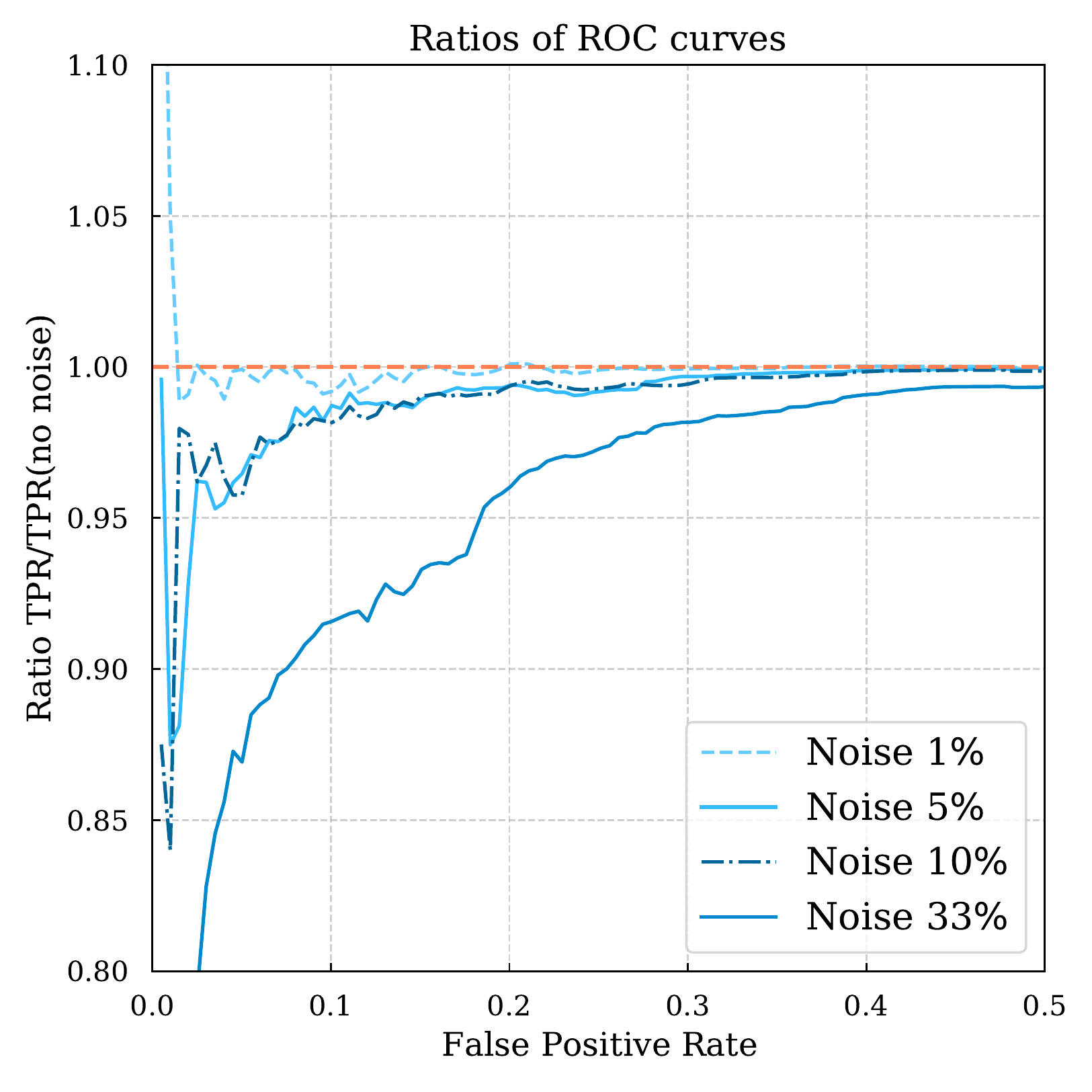}
\caption{Ratio of the ROC curves with different label noise levels to the one without noise. More specifically, we plot the ratio of the true positive rates, as a function of the false negative rate. The horizontal, orange, dashed, line corresponds to ratio = 1. Notice that the range in the false positive rate plotted is 0--0.5 since for higher values all curves converge to a value equal to one.}
\label{Fig: ROC_Ratio}
\end{figure}

\begin{table*}[!ht]
\caption{Classification metrics using the baseline split and different levels of random label noise (mislabeling) in the training set.}  
\label{table:Label_noise}
\centering
\begin{tabular}{|c||c|c|c|c|}
\hline
\diaghead{\theadfont Diag ColumnmnHead I}%
{\textbf{{\normalsize{Noise level}}}}{{\normalsize{\textbf{Metric}}}}& Accuracy& Completeness & Purity & AUC score \\
\hline \hline
\textbf{No noise} & 0.920 & 0.944 & 0.903& 0.974 \\
Noise 1$\%$ & 0.918 & 0.945& 0.899& 0.974 \\
Noise 5$\%$ & 0.914& 0.935 &0.890& 0.970\\
Noise 10$\%$ & 0.909& 0.897& 0.923& 0.969\\
Noise 33$\%$ & 0.883& 0.892& 0.880& 0.950\\
\hline
\end{tabular}
\end{table*}

A final source of uncertainty that we consider is the presence of mislabeled examples in the training set, also known as \textit{label noise}. In Sec.~\ref{sec: Interpretation} we showed that our dataset contains some LSBGs that are labeled as artifacts. Generally, we know that we were conservative when performing the visual inspection, meaning that we favored purity over completeness when assembling the LSBG catalog (if we were unsure if an object was a LSBG we preferred to flag it as an artifact).

Label noise has been extensively studied in the deep learning literature \citep[for overview papers see e.g.,][]{Frenay2014,Algan2019,Song2020}. The general conclusion is that neural networks are relatively robust to label noise, in some cases proving to perform well even in the presence of large noise \citep{Rolnick2017}.
Most of these studies consider well-known annotated datasets (MNIST, CIFAR, etc.) and introduce artificial label noise. 
However, it is interesting and useful to study potential effects of mislabeling in our dataset.

The best way to do this study would be to identify the mislabeled examples by carefully inspecting and reclassifying them. However, such a detailed study is time-intensive and beyond the scope of this work. Here we consider the impact of label noise in the following way: we randomly select a number of examples (LSBGs and artifacts alike) equal to $1\%,5\%,10\%$ and $33\%$ of images from the training set and we flip their label. We retrain the model each time and we evaluate on the test set. 

The classification performance metrics for each noise level (and for the baseline case without noise) can be found in Table \ref{table:Label_noise}. The corresponding ROC curves, that allow for a visual comparison of the performance of the models, can be seen in the right-hand panel of Fig.~\ref{fig: Uncertainties}. Furthermore, in Fig.~\ref{Fig: ROC_Ratio} we plot the ratios of the ROC curves for the models with different noise levels to that without label noise, for better inspection of the differences.
We can see that for small ($\lesssim 10\%$) label noise levels the reduction in performance is minimal.
Reduced accuracy only becomes noticeable once the label noise reaches $33\%$, though even in this case the accuracy reduction is not very large. Our results thus confirm the studies of \citet{Rolnick2017}.

So far we have considered a purely random noise, in the sense that we randomly selected an equal number of LSBGs and artifacts and flipped their labels. We repeated the above exercise introducing targeted noise---i.e., we selected only artifacts and changed their label to LSBGs in the first case, and LSBGs that changed their labels to artifacts in the second case. In both cases the results were qualitatively similar to the random noise case. 
We conclude that the presence of label noise in our sample, either random or biased against one of the two categories, does not significantly change the model performance. 

\section{Discussion and Conclusions}
\label{sec: Discussion_Conclusions}

In this work we presented the first application of deep learning, to the problem of automatic LSBG/artifact classification in astronomical images of LSBG candidates. 
This novel study was enabled by the availability of large samples of both LSBGs and artifacts from DES \citep{Tanoglidis2020}.

We showed that a simple CNN architecture with three convolutional and two dense layers can achieve classification accuracy of $92.0\%$ (completeness $94.4\%$ and purity $90.3\%$) and significantly improves over conventional machine learning models (SVMs and random forests) trained on \texttt{SourceExtractor}-derived features (accuracy $81.9\%$ and $79.7\%$, respectively).
This performance is found to be relatively robust to label noise.

We also demonstrated that knowledge obtained from one survey (training on DES data) can be transferred to another survey (prediction on HSC SSP data, accuracy $82.1\%$) and the performance of this \textit{transfer learning} can be significantly improved when the model is retrained on a small sample of examples from the new survey (accuracy $87.6\%$).

These results are promising and impactful for two reasons.
First, automating the classification process (or, at least, significantly reducing the need for visual inspection) will be necessary given the data volumes of future surveys such as LSST  and Euclid, and even future analyses of current surveys, such as the full 6-year of DES observations and future data releases from HSC SSP. 
Second, automated classification makes it much easier to characterize, in an unbiased way, the completeness/detection efficiency of future LSBG catalogs. The standard way to characterize detection efficiency is by injecting a large number of mock galaxies with a known parameters (e.g., effective radius, surface brightness, Sérsic index etc.) into the imaging data and then applying the same detection pipeline.
The efficiency can be calculated as a function of galaxy parameters by measuring the fraction of mock galaxies that are recovered \citep[e.g.,][]{Song2012,Suchyta:2016,Venhola2018}. 
To characterize the detection efficiency over the allowed space of galaxy parameters, it is often necessary to simulate more mock galaxies than are observed in the data itself. 
This makes unbiased human classification very challenging. 

We have presented the first study of CNNs for LSBG--artifact separation\footnote{Note that the reason we consider only these two categories is because of the preprocessing steps described in Sec.~\ref{sec: datasets}; because of them the final candidate sample is forced to contain only those two categories. Without this preprocessing one may consider a multi-class classification, like normal galaxies, stars.}. Our primary goal was to demonstrate the feasibility of such an approach, and we briefly outline possible further investigations.
Specifically, improvements in the CNN model, the training data, the use of domain adaptation techniques for transfer learning, and the systematic study of uncertainties would all improve future models for LSBG classification and discovery.

In particular, CNN architectures have a large number of tunable hyperparameters---e.g., number of layers, filters, kernel sizes, dropout levels, regularization parameters. Finding the optimal combination in a manner similar to that used for the SVM and random forest classifiers is computationally expensive. We have tested that the architecture presented here is robust to small changes -- e.g., the model with 3 convolutional layers performs better compared to one with 2 or 4 layers, etc. However, a grid of hyperparameters should be explored to ensure an architecture that gives the best results. This would likely require parallelizing the model training and evaluation processes to reduce computational time. Furthermore, more complex types of networks, such the Residual Neural Networks \citep[ResNet][]{ResNet} that allow for very long (large number of epochs) training, should be explored.

The quality of data used for training and testing is also very important. In Sec.~\ref{sec: mislabeling} we showed that the presence of label noise does not significantly change performance. However, the selection of objects for which the labels were flipped was totally random; in practice the objects that are more challenging to characterize and thus prone to mislabeling are of a specific type -- for example very faint or very compact. It would be useful to reclassify our sample into different confidence categories and check how the performance of the classifier changes when only high-confidence LSBGs and artifacts are used for training. Having a test set without label noise is also important; we have seen that some of the ``misclassifications" were actually result of label noise, thus leading to slightly misestimated performance metrics (here we refer to real label noise, not the artificial label we introduced in Sec.~\ref{sec: mislabeling}). Furthermore, \textit{data augmentation} is a commonly used technique that could be explored in the future \citep[e.g.,][]{Shorten2019}. Data augmentation is a regularization technique used to avoid overfitting, where one increases the number of training examples by adding slightly modified copies of the existing images (rotated, resized etc.).

The topic that likely requires the most detailed further study is that of transfer learning from one survey to another. Here our exploration was minimal, either applying the model trained on DES data directly to HSC SSP data or retraining the whole model on a very small set from HSC SSP.
More \textit{domain adaptation} techniques \citep[e.g.,][]{Kouw2018,Wang2018} (techniques that use algorithms trained in one or more ``source domains'' to a different, but related, ``target domain'') should be explored before choosing an approach to apply to forthcoming surveys (for domain adaptation applications in astronomy, see e.g., \citealt{Vilalta2019,Ciprijanovic2020b}). These techniques would allow the models to be successfully applied to the new data without the need to retrain the model later and more importantly to manually label new ``target" datasets since these techniques often use unlabeled target datasets. This makes the process much faster. Furthermore, the benefit of using larger example sets from the target survey for re-training of the model should also be explored.

Finally, the topic of uncertainty quantification in deep learning is a very active area of research; here a simple error estimation was presented. Future exploration should include bootstrapping the training set and not just the test set (something that would be computationally expensive and should be parallelized), sources of statistical and systematic errors (known as ``epistemic" and ``aleatoric" in the machine learning community) should be studied separately, as well as potential correlations between the two.

We plan to address some of these questions in future work, but the results presented here are already very promising for the upcoming analysis of the full six years of DES data. 
However, given the inherent challenges present in classifying very low-surface-brightness objects, the performance of a CNN classifier may never reach human-level accuracy. We argue that as we enter a new, big-data based, era in astronomy, the community should be ready to take a leap of faith in accepting the presence of small and well-characterized classification errors in favor of the great statistical power that comes when assembling large catalogs of objects that can illuminate the low-surface-brightness universe.

\section*{Acknowledgement}
We are grateful to Johnny Greco for providing us a list of visually rejected artifacts from the HSC survey, described in Sec \ref{sec: HSC} and to Brian Nord for useful comments and suggestions. 
A. \'Ciprijanovi\'c is partially supported by the High Velocity Artificial Intelligence grant as part of the Department of Energy High Energy Physics Computational HEP sessions program.

We acknowledge the Deep Skies Lab as a community of multi-domain experts and collaborators who've facilitated an environment of open discussion, idea-generation, and collaboration. This community was important for the development of this project.

This material is based upon work supported by the National Science Foundation under Grant No.\ AST-2006340.
This work was supported by the University of Chicago and the Department of Energy under section H.44 of Department of Energy Contract No.\ DE-AC02-07CH11359 awarded to Fermi Research Alliance, LLC. 
This work was partially funded by Fermilab LDRD 2018-052. 
This manuscript has been authored by Fermi Research Alliance, LLC under Contract No. DE-AC02-07CH11359 with the U.S. Department of Energy, Office of Science, Office of High Energy Physics. 

This project used public archival data from the Dark Energy Survey (DES). Funding for the DES Projects has been provided by the U.S. Department of Energy, the U.S. National Science Foundation, the Ministry of Science and Education of Spain, the Science and Technology FacilitiesCouncil of the United Kingdom, the Higher Education Funding Council for England, the National Center for Supercomputing Applications at the University of Illinois at Urbana-Champaign, the Kavli Institute of Cosmological Physics at the University of Chicago, the Center for Cosmology and Astro-Particle Physics at the Ohio State University, the Mitchell Institute for Fundamental Physics and Astronomy at Texas A\&M University, Financiadora de Estudos e Projetos, Funda{\c c}{\~a}o Carlos Chagas Filho de Amparo {\`a} Pesquisa do Estado do Rio de Janeiro, Conselho Nacional de Desenvolvimento Cient{\'i}fico e Tecnol{\'o}gico and the Minist{\'e}rio da Ci{\^e}ncia, Tecnologia e Inova{\c c}{\~a}o, the Deutsche Forschungsgemeinschaft, and the Collaborating Institutions in the Dark Energy Survey.

The Collaborating Institutions are Argonne National Laboratory, the University of California at Santa Cruz, the University of Cambridge, Centro de Investigaciones Energ{\'e}ticas, Medioambientales y Tecnol{\'o}gicas-Madrid, the University of Chicago, University College London, the DES-Brazil Consortium, the University of Edinburgh, the Eidgen{\"o}ssische Technische Hochschule (ETH) Z{\"u}rich,  Fermi National Accelerator Laboratory, the University of Illinois at Urbana-Champaign, the Institut de Ci{\`e}ncies de l'Espai (IEEC/CSIC), the Institut de F{\'i}sica d'Altes Energies, Lawrence Berkeley National Laboratory, the Ludwig-Maximilians Universit{\"a}t M{\"u}nchen and the associated Excellence Cluster Universe, the University of Michigan, the National Optical Astronomy Observatory, the University of Nottingham, The Ohio State University, the OzDES Membership Consortium, the University of Pennsylvania, the University of Portsmouth, SLAC National Accelerator Laboratory, Stanford University, the University of Sussex, and Texas A\&M University.

Based in part on observations at Cerro Tololo Inter-American Observatory, National Optical Astronomy Observatory, which is operated by the Association of Universities for Research in Astronomy (AURA) under a cooperative agreement with the National Science Foundation.

\appendix

\section{Three-class classification}
\label{sec: three_class}

\begin{figure}[!ht]
\centering
\includegraphics[width=0.9\columnwidth]{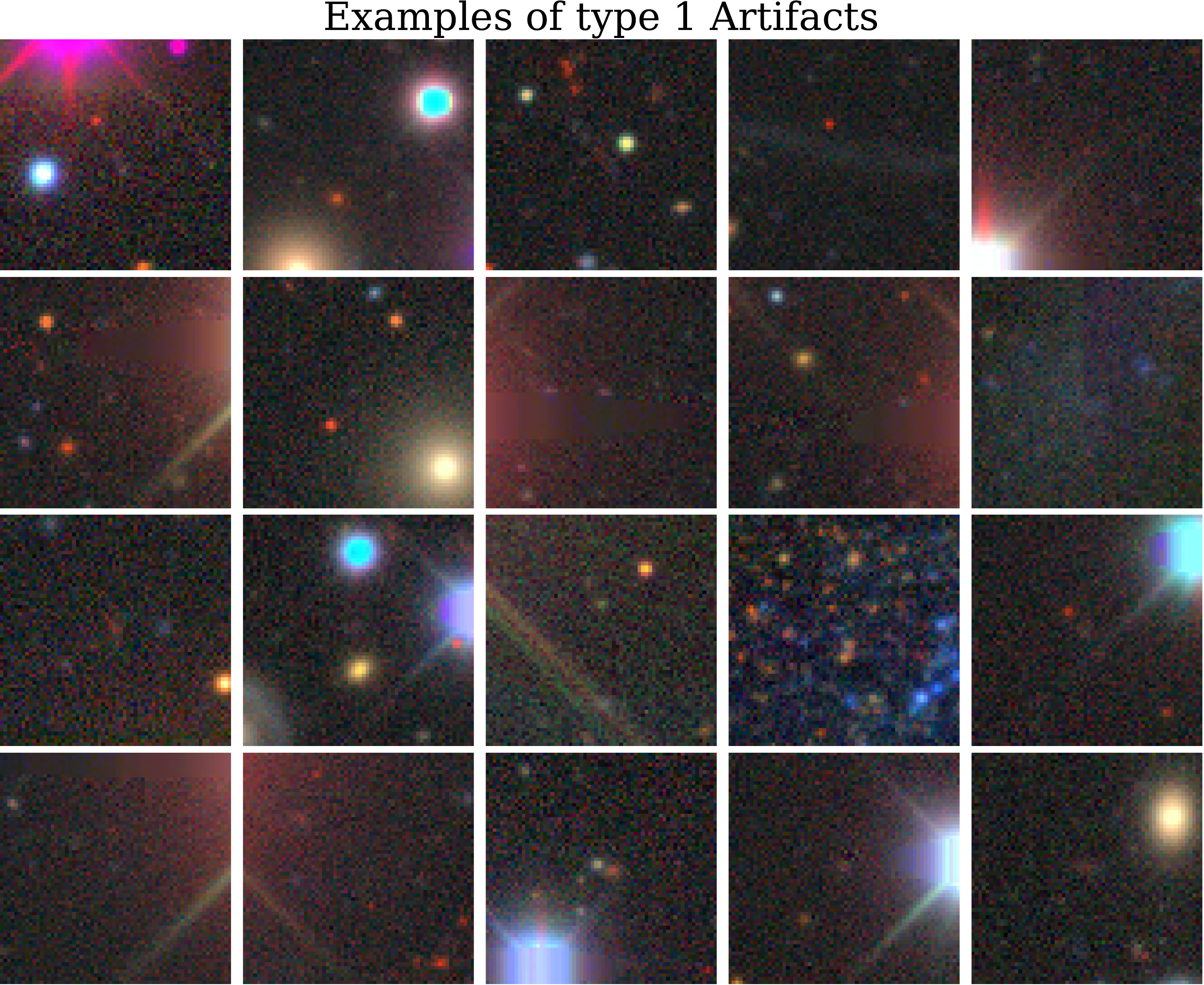}
\caption{Randomly selected examples of type 1 artifacts (see text).}
\label{Fig: Artifacts_1}
\end{figure}

 \begin{figure*}[!h]
\centering
\subfigure[]{\includegraphics[width=0.47\textwidth]{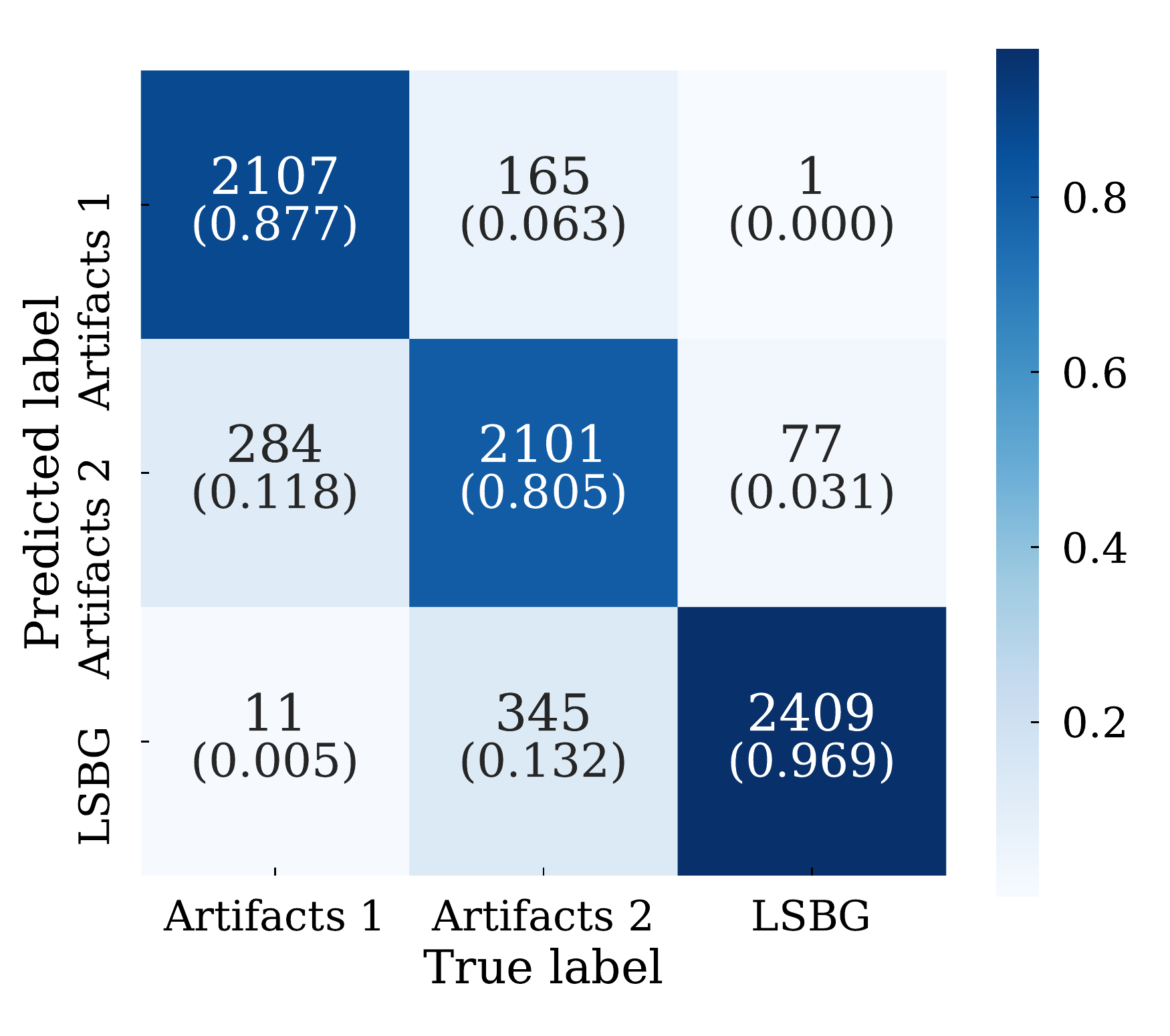}}
\hspace*{\fill}
\subfigure[]{\includegraphics[width=0.47\textwidth]{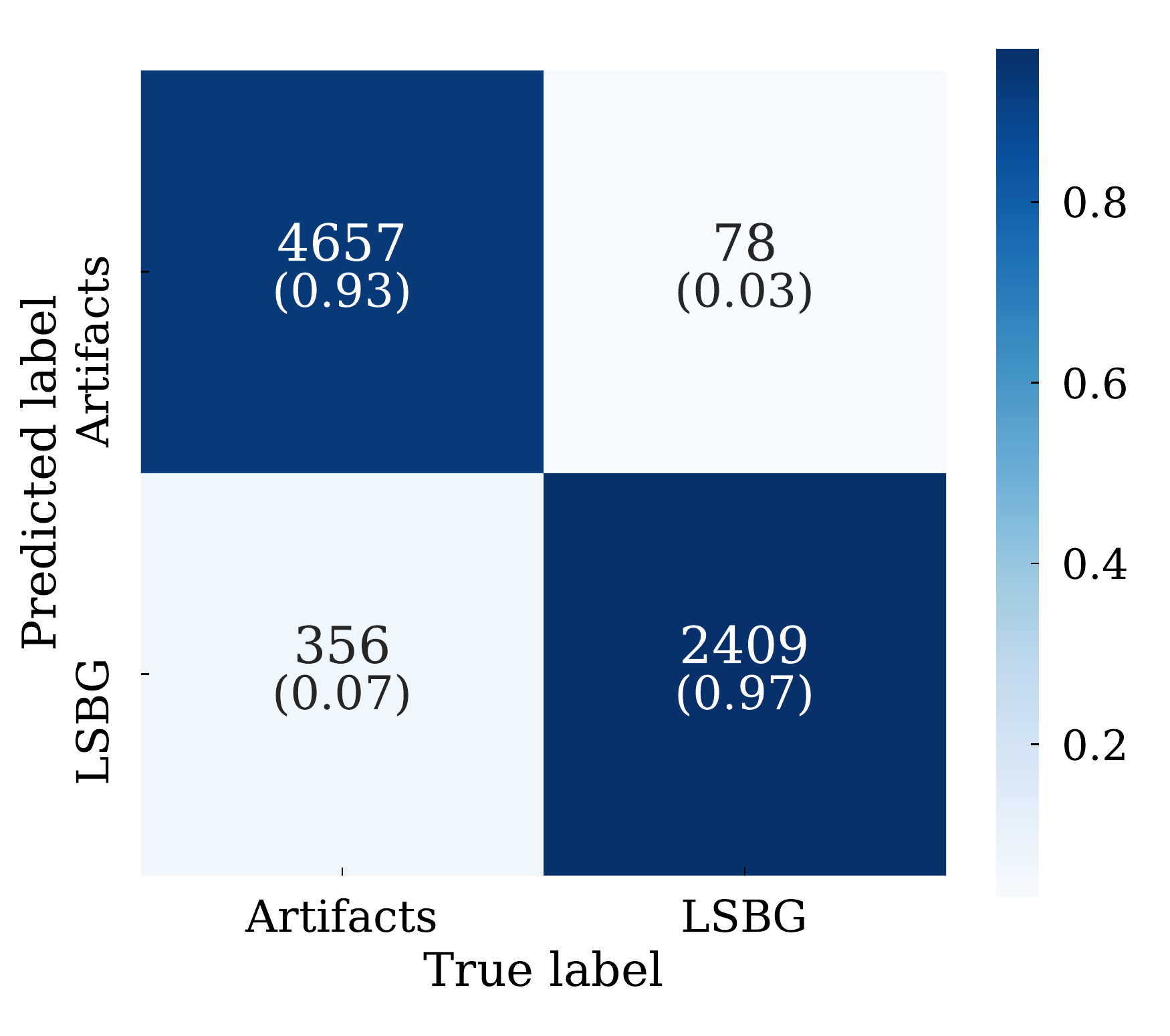}}
\caption{(a) Confusion matrix for the three-class classification problem, where two artifact classes are considered. The numbers in parentheses correspond to the normalized values. (b) ``reduced" confusion matrix, where the two artifact classes have been combined.}
\label{fig: Three_class}
\end{figure*}

In the main body of this paper we considered a two-class classification: one (positive) category of LSBGs and one (negative) category of artifacts -- those visually rejected in the third step of the LSBG catalog generation in \citet{Tanoglidis2020}, also described in Sec.~\ref{sec: LSB_and_Art}. These artifacts were the hardest to classify, since they had been classified as LSBGs by an SVM classifier trained on \texttt{SourceExtractor} features. Before the  visual inspection the same classifier had rejected a very large number of artifacts (most of them correctly).

We also investigated the performance of \textit{DeepShadows} on a three-class classification problem where two artifact classes are considered, by adding a sample of 20,000 randomly selected artifacts from those rejected by the SVM classifier. 
We call this class ``Artifacts 1", while the artifacts considered in the main body of the paper are called ``Artifacts 2". 
In Fig.~\ref{Fig: Artifacts_1} we present a randomly selected sample of artifacts of the first kind. We see that these artifacts are dominated by the presence of strong diffraction spikes, and they are less confusing (more easily recognized as artifacts) than those presented in Fig.~\ref{fig:Cutout_Examples}.

We keep the same architecture for \textit{DeepShadows}, except that the last dense layer has size of three. We also change the final activation function to softmax and the loss to categorical crossentropy. We train again the model for 100 epochs with a batch size of 64. We split the total dataset of 60,000 object into 45,000 (training), 7,500 (validation), and 7,500 (test) sets.

The resulting three-class confusion matrix of the predictions on the test set can be seen on the left-hand side of Fig.~\ref{fig: Three_class}. We can see that there is very small confusion between the ``Artifacts 1" class and ``LSBGs" categories, confirming our notion that these are artifacts that can be very easily excluded. Interestingly, the classifier is able to distinguish between the two categories of artifacts, too (with some confusion, of course, accuracy $\sim 90\%$ when calculating for the submatrix between artifacts only).

In the right-hand panel of Fig.~\ref{fig: Three_class} we combine the two artifact categories, in order to better see the confusion between artifacts and LSBGs. Note that this is now an imbalanced two-class problem, since the artifacts class is twice as large as the LSBGs class. For that reason, accuracy is not a good metric but we can still calculate the completeness and purity, which are more meaningful metrics for the problem at hand. These numbers are $96.8\%$ and $87.1\%$, respectively and they are comparable (slightly less in purity) to those from the 2-class model discussed in the text. 

Our conclusion is that there is not much benefit in using a three-class classification, unless we prefer to eliminate the SVM classification step in future applications. 
 
 \section{Grad-CAM details}
 \label{sec: Grad_CAM}

We present here some technical details of the Grad-CAM technique for highlighting the most important regions for classification in an image. More details can be found in the original article \citep{GRADCAM}. We also suggest the following blog post\footnote{\url{https://fairyonice.github.io/Grad-CAM-with-keras-vis.html}} for an explanation of the technique and its application using \texttt{Keras}.

We define $A^k$ to be the $k-$th $(k=1,\dots,K)$ feature map of the last convolutional layer, that has dimensions $m \times n=Z$ (Pixel values $A_{ij}^k$, $i=1,\dots,m$, $j=1,\dots,n$). Let also $y^c$ the output score (probability) for the class $c$ (obviously, here we have only one positive class, thus only one probability score).

If each convolutional kernel captures a specific visual pattern, then each feature map of the final layer will show where this visual pattern exists in the image. We can thus imagine that the classification output depends on a weighted sum of all the feature maps, with weights depending on the importance each feature has for class $c$. So, the Grad-CAM maps can be written as: $L^c_{\text{\scriptsize{Grad-CAM}}} \sim \sum_k \alpha_k^c A^k$.

What are the class-dependent weights $a_k^c$? The idea is that the gradients of the output score with respect to the $(i,j)$ pixel of the $k-$th feature map, $\partial y^c/\partial A_{ij}^k$, measures the effect of that pixel to the classification score. Grad-CAM then proposes to take the average of all pixels (also known as average pooling) as that the weight for map $k$ and class $c$ is:
\begin{equation}
    a_k^c = \frac{1}{Z} \sum_{i=1}^m \sum_{j=1}^n \frac{\partial y^c}{\partial A_{ij}^k}.
\end{equation}
Finally, a ReLU function is applied to the weighted sum, to keep the positive regions, so we get:
\begin{equation}
  L^c_{\text{\scriptsize{Grad-CAM}}} = \text{ReLU}\left( \sum_{k=1}^K a_k^c A^k\right), 
\end{equation}
which produces a localization map that retains the spatial information present in the last convolutional layer.

\bibliographystyle{model2-names}
\bibliography{main}







\end{document}